\documentclass[10pt,journal]{IEEEtran}

\ifCLASSOPTIONcompsoc
  \usepackage[nocompress]{cite}
\else
  \usepackage{cite}
\fi

\usepackage{times}  %Required
\usepackage{helvet}  %Required
\usepackage{courier}  %Required
\usepackage{url}  %Required
\usepackage{graphicx}  %Required
\usepackage{comment}
\usepackage{booktabs} % For formal tables
\usepackage{threeparttable}
\usepackage{tabularx}
\usepackage{multirow}
\usepackage{hyperref}
\usepackage{amsmath}
\usepackage{physics}
\usepackage{subfig}
\usepackage{slashbox}
\usepackage{todonotes}
\usepackage{multirow}
\usepackage{soul}
\usepackage[multiple]{footmisc}

\newcolumntype{L}[1]{>{\raggedright\let\newline\\\arraybackslash\hspace{0pt}}m{#1}}
\newcolumntype{C}[1]{>{\centering\let\newline\\\arraybackslash\hspace{0pt}}m{#1}}
\newcolumntype{R}[1]{>{\raggedleft\let\newline\\\arraybackslash\hspace{0pt}}m{#1}}

\makeatletter
\def\ps@IEEEtitlepagestyle{
  \def\@oddfoot{\mycopyrightnotice}
  \def\@evenfoot{}
  \def\@oddhead{\mycitationnotice}
  \def\@evenhead{}
}

\def\mycopyrightnotice{
  {\footnotesize
  \begin{minipage}{\textwidth}
  \centering
  Copyright~\copyright~2019 IEEE. Personal use of this material is permitted. Permission from IEEE must be obtained for all other uses, in any current or future media, including reprinting/republishing this material for advertising or promotional purposes, creating new collective works, for resale or redistribution to servers or lists, or reuse of any copyrighted component of this work in other works. DOI: \href{https://doi.org/10.1109/TCSS.2019.2943238}{10.1109/TCSS.2019.2943238}.
  \end{minipage}
  }
}

\def\mycitationnotice{
  {\footnotesize
  \begin{minipage}{\textwidth}
  \textcolor{blue}{This is a preprint version. Please cite the work by using the reference on \href{https://ieeexplore.ieee.org/document/8863642}{https://ieeexplore.ieee.org/document/8863642}\\
R. Sequeira, A. Gayen, N. Ganguly, S. K. Dandapat and J. Chandra, "A Large-Scale Study of the Twitter Follower Network to Characterize the Spread of Prescription Drug Abuse Tweets," in IEEE Transactions on Computational Social Systems, vol. 6, no. 6, pp. 1232-1244, Dec. 2019, doi: 10.1109/TCSS.2019.2943238.}
  \end{minipage}
  }
}

\begin{document}
\title{A Large Scale Study of the Twitter Follower Network to Characterize the Spread of Prescription Drug Abuse Tweets}

 \author{Ryan Sequeira, 
 Avijit Gayen, 
 Niloy Ganguly, 
 Sourav Kumar Dandapat, 
 Joydeep Chandra\\
 
% }
        
%\IEEEcompsocitemizethanks{\IEEEcompsocthanksitem Prof. Niloy Ganguly is with the Department of Computer Science \& Engineering, Indian Institute of Technology Kharagpur,
%India.\protect\\
%
%E-mail: niloy@cse.iitkgp.ernet.in
%\IEEEcompsocthanksitem Ryan Sequeira, Avijit Gayen, Dr. Sourav Kumar Dandapat and Dr. Joydeep Chandra are with the Department of Computer Science \& Engineering, Indian Institute of Technology Patna, India. \protect\\
%E-mail: \{sequeira.mtcs16, avijit.pcs13, sourav, joydeep\}@iitp.ac.in}

%\thanks{Manuscript received April 05, 2019; revised July 31, 2019; accepted September 11, 2019.}
}

% The paper headers
%\markboth{IEEE Transactions on Computational Social Systems}%
%{A Large Scale Study of the Twitter Follower Network to Characterize the Spread of Prescription Drug Abuse Tweets}

\IEEEtitleabstractindextext{%
\begin{abstract}
In this paper, we perform a large-scale study of the Twitter follower network, involving around $0.42$ million users who justify drug abuse, to characterize the spreading of drug abuse tweets across the network.  Our observations reveal the existence of a very large giant component involving $99\%$ of these users  with dense local connectivity that facilitates the spreading of such messages. We further identify active cascades over the network and observe that cascades of drug abuse tweets get spread over a long distance through the engagement of several closely connected groups of users. Moreover, our observations also reveal a collective phenomenon, involving a large set of active fringe nodes (with a small number of follower and following) along with a small set of well-connected non-fringe nodes that work together towards such spread, thus potentially complicating the process of arresting such cascades. 
Further, we discovered that the
engagement of the users with respect to certain drugs like Vicodin, Percocet and OxyContin, that were observed to be most mentioned in Twitter, is instantaneous. 
On the other hand for drugs like Lortab, that found lesser mentions, the engagement probability becomes high with increasing exposure to such tweets, thereby indicating that drug abusers engaged on Twitter remain vulnerable to adopting newer drugs, aggravating the problem further. 
 
\end{abstract}

}

\maketitle
%\citationnotice
%\copyrightnotice

\IEEEdisplaynontitleabstractindextext

\IEEEpeerreviewmaketitle

\section{Introduction}
\label{sec:intro}

\IEEEPARstart{T}{he} enormous popularity of social media like Twitter makes it suitable as an advertisement platform for promoting drug-abuse.
Keeping in view the spread and impact of drug-abuse (there is an estimated count of $190,000$ premature drug-related deaths~\footnote{https://www.cdc.gov/nchs/data/health\_policy/monthly-drug-overdose-death-estimates.pdf}), and the limitations of traditional \emph{``prescription drug monitoring programs''}(PDMPs) in understating the severity of this issue~\cite{fink2018association} there is a need to have a deeper understanding of how social media is playing a role in promoting the drug menace. 

\par We focus our attention on the Twitter platform, that is one of the key media used to spread information related to drugs. Several background works exist that have highlighted the role of Twitter in the sale of illicit drugs~\cite{mackeyOnlineSale17} and the promotion of drug-abuse~\cite{mengplos17}.  Possible surveillance strategies for identifying such retailers and drug-abusers have also been well explored~\cite{kalyanamSurveillanceSurvey2017}. However, an important aspect that needs to be carefully investigated is the networked effect of Twitter that may amplify the spread of drug-abuse tweets among users. Preliminary studies of Twitter users  in~\cite{hanson2013exploration}  reveal the existence of drug-related social circles (densely connected neighbor set of a user) around certain active users who tweet frequently, mentioning the different effects that specific drugs produce.  
Although it is not completely clear whether such active circles can influence non drug-abusers towards abuse, however, from concepts like the ``{\em uses and gratification theory}'' and ``\emph{Health Communication Media Choice}'' (HCMC) model ~\cite{duffyEmerging09}, it can be argued that such discussions that glorify drug-abuse are likely to engage users who justify drug-abuse in further deliberations to satisfy their communication needs. Consequently, cascades of user engagement, if formed  through such discussions, would volume up as social advertisements that would downplay the ill effects of drug-abuse and may influence vulnerable users towards such practice. 
In this paper, we discover and characterize such cascades over the Twitter follower network, where users participate in discussions that treat drug-abuse positively.

To identify these cascades, we propose a technique that uncovers around $0.42$ million unique users who were engaged in either self-reporting or promoting prescription drug-abuse through tweets. The enormity of this number reflects the huge role being played by social media in promoting prescription drug-abuse on Twitter. However, the dynamics of these cascades are mainly driven by the users' engagement behavior, as well as the underlying structure of the follower network. Hence in our study, we follow a principled approach by first investigating the underlying follower network of these unique users and subsequently the characteristics of these users in terms of their engagement behavior and positional importance in the network, before finally studying the key features of the cascades. The {\em follower} relation among these users is used to create a network with directed edges. 
Investigation reveals that this network is almost entirely connected with around $91.37\%$ of the nodes being in the largest strongly connected component.
We also observe very high reciprocity of the links in the network.
All these network properties indicate the network structure is amenable to the large-scale spread of information. 
The spread, however, depends on the activeness of the users (frequency of engagement) as well as their position in the network. We assessed the activeness of the users and found that around one-fourth of the users are active. 
Moreover, the reach of these active nodes is considerable, as they cover a substantial part of the network.

\par From the detailed study of the nature of the network and its participating users, it is clear that a structure susceptible to facilitate the formation of cascades exist.  We discover around  $50,409$ cascades, some of them with sizes reaching to thousands and extending over several hops. The network study of such cascades reveals high structural virality, i.e. the cascades are not driven by a single {\em important} node, which in turn makes them difficult to control. Further, the network among the nodes participating in each cascade exhibit significantly high clustering coefficient and reciprocity in their follower relationships. All these observations provide a strong evidence that drug-abuse tweets traverse through several groups of closely connected users. Moreover, the study of the characteristics of these users reveals equally important contributions, irrespective of their position and activeness, in the spreading process. 
The phenomenon indicates that organic collaboration among  a large set of  nodes is largely responsible for the emergence of  a cascade; thus it may be difficult to control the cascades by  eliminating a few targeted nodes. Finally, guided by the  metrics associated with social contagion processes~\cite{RomeroPersistenceWWW2011} (discussed in detail later),  we find that users engage (through tweets or retweets) differently when exposed to drug-abuse tweets with different drug names. On exposure to tweets with highly mentioned drug names like Vicodin, Percocet and OxyContin, the engagement is instantaneous, whereas for less popular drugs like Lortab the chances of engagement increase with increasing exposure to such mentions. This signifies that drug-abusers engaged on Twitter remain at risk of adopting newer drug types to which they are continually exposed through Twitter discussions.

\par The rest of the paper is organized as follows. In the next section, we highlight the related works. In section~\ref{sec:data}, we describe the detailed dataset used in our work. In this section, we explain the data collection process as well as the classification method to identify the drug-abuse tweets. In section~\ref{sec:PDAN}, we describe the follower network formation method and further detail its network characteristics. We subsequently discuss the user characteristics in section~\ref{sec:user-character}. The details about the spreading pattern of the tweets across the follower network are outlined in section~\ref{sec:spread}. Finally, we summarize our findings in section~\ref{sec:conclusion}.
\section{Related Work}
\label{sec:related}
 A plethora of recent work uses social media to gather information and in turn, provide solutions to various issues related to health.  
For example, social media has played an vital role in providing rich information for inferring mental health conditions (especially depression~\cite{munmumDepressionICWSM13}, mood instabilities~\cite{ChoudhuryMoodHCIK15}, suicidal risks~\cite{choudhurySuicideICWSM17} and the effects of psychiatric medication~\cite{saha2019social}), as well as lifestyle-related conditions like overeating, alcoholism and 
smoking~\cite{tamersoySmokingDH17}.
Technological approaches are being leveraged for addressing critical issues like the early prediction of such diseases~\cite{munmumDepressionICWSM13}, increased support and service engagement~\cite{rubyaSubstanceSupportCSCW2017} and a decrease in the duration of untreated disorders~\cite{ChoudhurySelfDisclosureICWSM14}. These works provide a direction to the critical issues, concerning the psychological problems (including the drug-abuse problem), that require immediate attention.

\par Recently, prescription drug-abuse is receiving increasing attention due to its significant spread and the casualties involved~\cite{kalyanam2017exploring,klein2017detecting}. Majority of these work are directed towards identifying content on social media that reveal drug-abuse behavior of the users. Such techniques include both supervised classification ~\cite{sarker2016social,buntain2015your,hu2018deep, mahata2018detecting} as well as unsupervised methods~\cite{kalyanam2017exploring,ding2016analyzing} for identifying drug-abuse tweets from the tweet stream.
Complementary to the problem of drug-abuse detection, works based on identifying alternative opioid recovery treatments on social media also exist\cite{chancellor2019discovering}.
Such data can be leveraged to gain insights about the microscopic behavior of the corresponding users and their role in spreading drug-abuse tweets in the network, although only a few works have attempted to do so~\cite{mackeyOnlineSale17,hanson2013exploration}. One of the primary goals of this paper is to work towards these objectives.
\par In~\cite{hanson2013exploration}, the authors pointed out the influence of neighbors (network effect) on the participation of users in discussions related to drug-abuse. They observed the presence of social circles in which active users (users who frequently discuss the abuse of specific drugs on Twitter) are likely to be surrounded by users who also participate in similar discussions, exhibiting a high content correlation among them. While this work provides preliminary insights about the tweeting behavior among these participating users, revealing the existence of a possible group phenomenon; to the best of our knowledge, no other work has attempted to take a closer look
into the spread of drug-abuse discussions on social media platforms. However, the study of the propagation of different tweet content to understand human behavior has been a focus area across various topical domains.
One of the earlier works on information flow on Twitter~\cite{wuWWW2011} showed the presence of a few elite users in Twitter who generate a majority of the content that is consumed by ordinary users. Several other works have also investigated the user characteristics and the role of influential users in information propagation across social networks~\cite{aralScience2012,bakshyInfluencerWSDM2011}. Subsequent empirical works have examined several other factors like the underlying network of the users~\cite{bakshySocialWWW2012,chengCascadePredictionWWW14,wengDiffusionNature13}, the user characteristics~\cite{romeroPassivity11,saezTrendsetterKDD12,gomezQuantifyingICWSM14} the role of  content~\cite{szaboContentCACM10,tanWordingACL14} and even the role of the underlying diffusion protocols~\cite{cheng2018diffusion} in such propagation. Selected works have investigated the effects of both users and content in information propagation~\cite{hoangTrackingCIKM016,rudraIdiomsPAKDD15}.
Since the dynamics of information propagation across networks vary with topics and content, this motivates the need to investigate the spreading behavior in the context of drug-abuse tweets by inspecting the cascades of user engagement, the role of the network and the user characteristics in the spreading process.

\par Spreading of various social behaviors has been investigated in several works, like the cessation of smoking~\cite{christakisSmoking2008}, online sharing~\cite{spragueContagionPLOS2017}, and political controversies~\cite{RomeroPersistenceWWW2011}. These works highlight the importance of collective dynamics, often modeled through a complex contagion phenomenon, in the spreading of social behavior. As these works can eventually help in controlling viral spread when such propagation is not desired (like in case of drug-abuse tweets), there is a need to look into the generation of drug-abuse tweets through the prism of such models. As there is still a wide gap in understanding the characteristics of the social network through which such drug-abuse tweets spread and the role of the users that influence {\em spreading} of drug-abuse content, we believe this paper would contribute in filling this gap. In the next section, we describe the dataset used for this study.
\section{Dataset}
\label{sec:data}
\begin{figure*}[ht]
\centering
\includegraphics[width=\textwidth]{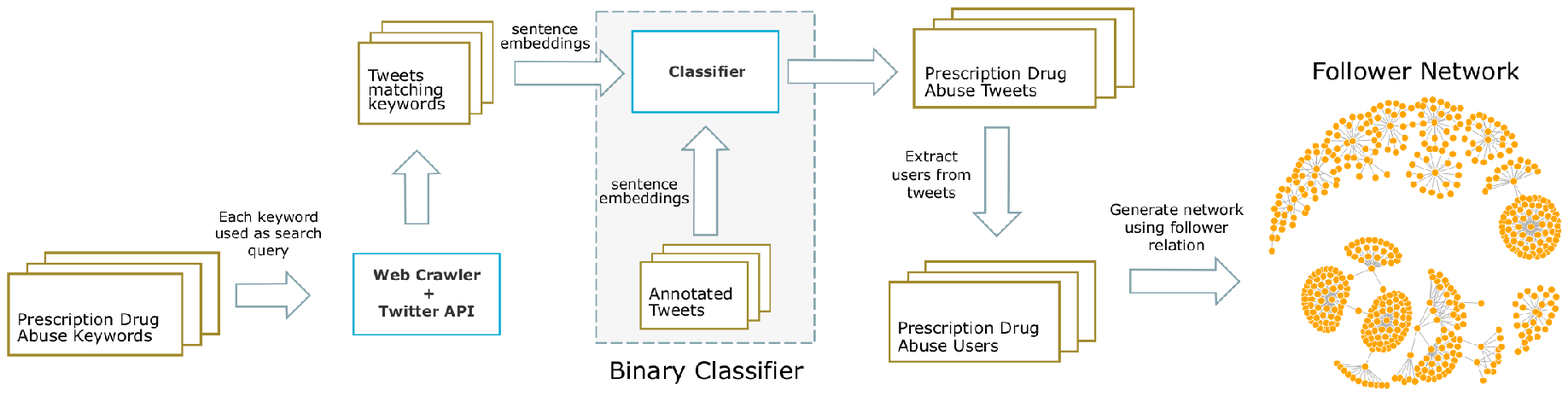}
\caption{Steps of follower network formation.}
\label{fig:method_steps}
\end{figure*}

\begin{table}
	\caption{List of generic and brand names of prescription opioids medically used to treat pain\textsuperscript{\ref{note1}}.}
    \label{tab:keywords}
\centering
\footnotesize
\begin{tabular}{| c | c |}
    	\hline
        \textbf{Generic names} & \textbf{Brand names} \\
        \hline
        oxycodone & OxyContin, Percodan, Percocet \\
        hydrocodone & Vicodin, Lortab, Lorcet \\
        diphenoxylate & Lomotil \\
        morphine & Kadian, Avinza, MS Contin \\
        codeine & - \\
        fentanyl & Duragesic \\
        propoxyphene & Darvon \\
		hydromorphone & Dilaudid \\
        meperidine & Demerol \\
        methadone & - \\
        \hline
	\end{tabular}
\end{table}

In this section, we provide a detailed description of the Twitter dataset along with the data collection methodology and the preprocessing techniques used. We subsequently describe the classification technique used to identify tweets that are promoting or reporting prescription drug-abuse. In line with the literature~\cite{hanson2013exploration}, the corresponding users engaged in such tweets are henceforth termed as {\em drug-abusers}. Based on the follower relation among these drug-abusers, a network is created. The steps followed to form the network is pictorially represented in figure~\ref{fig:method_steps}.

\subsection{Data Collection} The data collection steps can be briefly described as follows:
\begin{enumerate}
\item We prepared a set of drugs names (see table \ref{tab:keywords}) that have been marked and listed for abusive use in the past by the National Institute on Drug Abuse (NIDA)\footnote{\label{note1}https://teens.drugabuse.gov/drug-facts/prescription-pain-medications-opioids}.
The generic and brand names of these drugs were used as search \emph{keywords} for collecting the drug-related tweets using the web-based crawler, ``\textit{Get Old Tweets}''. This provided all the searchable tweets from January 2012 to July 2017 containing the drug names.
\item As only limited information about the tweets was being provided by the web-based crawler, we used the ``\textit{tweet id}'' of the returned tweets to further query and extract the complete meta-data using the Twitter API\footnote{https://api.twitter.com/1.1/statuses/show.json}. 
\item As retweets could not be retrieved using this web-based crawler, they were obtained using a different Twitter API\footnote{https://api.twitter.com/1.1/statuses/retweets/:id.json}. 
\item Finally, non-English tweets were identified using LangID\footnote{https://github.com/saffsd/langid.py} and discarded.

\end{enumerate}

Using this approach, we collected more than $2$ million drug-related tweets.  However, we observed that this collected tweet set included both kinds of tweets: those promoting drugs or reporting drug-abuse as well as those spreading awareness or rehabilitation and treatment information, that we term as non-abuse tweets. Hence, we applied several machine learning techniques to identify drug-abuse tweets, which is detailed in the next section. 

\subsection{Classification of Drug-Abuse Tweets}

\begin{table}[!hbt]
\caption{Performance of binary classification of prescription drug-abuse tweets. 10-fold cross-validation is used to measure the $F_1$ score of the two classes (Drug-Abuse and No-Abuse) and the overall accuracy of the classifier.}
\label{tab:classifier-results}
\centering
\begin{tabular}{ |l|C{1.2cm}|C{1.2cm}|c| } 
 \hline 
 \textbf{Classifier} & \textbf{DA $F_1$}  & \textbf{NA $F_1$}  & \textbf{Accuracy} \\
  \hline
  % n-gram features
  \multicolumn{4}{|c|}{N-grams as features and handcrafted features} \\
  \hline
  Naive Bayes & $0.752$ & $0.701$ & $72.87\%$ \\
  SVM & $0.787$ & $0.759$ & $77.42\%$ \\
  Random Forest & $0.842$ & $0.794$ & $82.12\%$ \\
  Logistic Regression & $0.701$ & $0.694$ & $69.75\%$ \\
  AdaBoost & $0.740$ & $0.613$ & $68.91\%$ \\
  XGBoost & $0.776$ & $0.694$ & $74.13\%$  \\
  Bagging & $0.770$ & $0.752$ & $76.14\%$  \\
  Voting & $0.786$ & $0.758$ & $77.23\%$  \\
  \hline
  
  % sentence embeddings
  \multicolumn{4}{|c|}{Sentence embeddings and handcrafted features} \\
  \hline
  Naive Bayes & $0.758$ & $0.735$ & $74.71\%$ \\
  \textbf{SVM} & \textbf{$0.857$} & \textbf{$0.846$} &\textbf{$85.16\%$} \\
  Random Forest & $0.814$ & $0.806$ & $81.03\%$ \\
  Logistic Regression & $0.811$ & $0.803$ & $80.70\%$ \\
  AdaBoost & $0.782$ & $0.770$ & $77.61\%$  \\
  XGBoost & $0.817$ & $0.810$ & $81.34\%$  \\
  Bagging & $0.795$ & $0.780$ & $78.85\%$  \\
  Voting & $0.843$ & $0.832$ & $83.83\%$  \\
  \hline
  
  % deep learning
  \multicolumn{4}{|c|}{Deep learning with word embeddings} \\
  \hline
  LSTM & $0.807$ & $0.812$ & $81.03\%$ \\
  RNN & $0.819$ & $0.820$ & $81.99\%$ \\
  RCNN~\cite{lai2015recurrent} & $0.805$ & $0.812$ & $80.90\%$ \\
  TextCNN~\cite{kim2014convolutional} & $0.830$ & $0.837$ & $83.38\%$\\
  \hline
   
\end{tabular}
\end{table}

\begin{table*}[t]
 \caption{Example of the variety of tweets that match our keywords. The keywords (listed in table \ref{tab:keywords}) that are used to search prescription drug-abuse tweets are highlighted. All the tweets in this table are paraphrased to maintain the anonymity of the users.}
    \label{tab:tweet_examples}
    \centering
    \scriptsize
    \renewcommand{\arraystretch}{1.5}
    \begin{tabular}[t]{|p{0.13\textwidth}|p{0.32\textwidth}|p{0.13\textwidth}|p{0.32\textwidth}|}
        \hline
        \multicolumn{2}{|c|}{\textbf{Drug-Abuse Examples}} &  \multicolumn{2}{c|}{\textbf{Non-Abuse Examples}} \\
        \hline
        \textbf{Category} & \textbf{Tweet} & \textbf{Category} & \textbf{Tweet}\\
        \hline
        \textbf{Addiction} & I was an addict, a complete addict! I was consuming more than \textit{\textless NUMBER\textgreater} mg of \textbf{Morphine} and \textbf{OxyContin} each day. & \textbf{Metaphor / Sarcasm / Jokes} & What you call Alabama Shakes we call that \textbf{OxyContin} withdrawal in the state of Ohio. \\ 
        
        & I am officially addicted to \textbf{Vicodin}... I need help. &  & My grandmother was telling about how much she money could earn by selling her \textbf{Percocet} on the streets. lmfao. \\ 
        
        \hline
        
        \textbf{Co-ingestion} & Washing \textbf{Demerol} down my throat with some vodka.  I sense that I have officially failed at life. & \textbf{Awareness} & Left over \textbf{Vicodin}: Flush or Trash? Link to FDA for Safe Medication Disposal.   \textit{\textless URL\textgreater}. A useful homepage link for patients? \\ 
        
        & \textbf{Vicodin} with Weed = a long sleep. & & The CDC says \textit{\textless NUMBER\textgreater} people died this year from prescription opioid overdose. \#\textbf{Percocet} \\ 
        
        \hline

        \textbf{Alternate modes of ingestion} & I'm sitting on a large leather chair railing a ton of \textbf{Dilaudid}. & \textbf{Rehabilitation} & \textbf{Vicodin} Rehab in Florida, Florida Center for Recovery \textit{\textless URL\textgreater}. \\ 
        
         & I'm about to snort some of the \textbf{OxyContin} off my table \#YOLO &  &  Behind the Dependence on \textbf{OxyContin} and Transition to Heroin @ Trusted Heroin Rehab \textit{\textless URL\textgreater}\\ 
        
        \hline
        
        \textbf{Recreation} & Have I at any point referenced how much I recreationally enjoy \textbf{Vicodin}? & \textbf{Pop-culture references} & \textbf{OxyContin}, Xanax bars, \textbf{Percocet} and \textbf{Lortab} / Valiums, Morphine patches, ecstasy / It's all up for grabs \\ 
        
        & Having a \textbf{Vicodin} party at my place, who wants to join? \#insomnia & & Don't shoot, you can't fight a viking on \textbf{Vicodin}. Can you?  \\ 
        
        \hline

        \textbf{Selling} & I'm amassing all my remaining \textbf{Vicodin} and \textbf{Percocet} from my previous two medical procedures and  selling them as soon as I get back to the burgh & \textbf{News} & Washington city devastated by \textbf{OxyContin} addiction sues Purdue Pharma — claiming drugmaker put profits over citizens \textit{\textless URL\textgreater}\\ 
        
        & Giving out \textbf{Percocet}, \$\textit{\textless NUMBER\textgreater} a pill... Get your hands on them while they last!! & & Heroin makes lethal comeback after \textbf{OxyContin} becomes more difficult to crush - Alaska Dispatch \textit{\textless URL\textgreater}\\ 
        
        \hline
        
        \textbf{Ingestion} & One of the metrits is that I get to pop \textbf{Vicodin} like breath mints. So that's always a good thing. & \textbf{Medical treatment} & Consumed a \textbf{Vicodin} to get rid of the pain in my mouth only to throw out everything a few hours after my surgery.\\

        & High as a kite after taking that \textbf{Vicodin}. & & Surgery went well, and i'm happy that it's over. Its time to take some \textbf{Vicodin}!! \\
        
        \hline
        
        \textbf{Illegal online pharmacies} & Order High quality \textbf{OxyContin} Online, \textit{\textless NUMBER\textgreater}\% discount. No Prescription Required \textit{\textless URL\textgreater} & \textbf{Pain} & Sadly, the \textbf{Percocet} isn't helping me with my knee pain. I may need to explore something different. \\ 
       
        & Did someone say \textbf{Percocet}? Buy \textbf{Percocet} Online \hspace{0.5cm} \textit{\textless URL\textgreater} & & My knee is in great pain. I need to take a \textbf{Percocet} for the pain. \\ 
        
        \hline
    \end{tabular}
\end{table*}

One of the important requirements in identifying the drug-abusers is to classify the tweets based on whether they promote (or self-report) prescription drug-abuse or not. As a significant proportion of the tweets are meant towards increasing awareness against drug-abuse, we need to filter out tweets that promote or self-report drug-abuse from the rest of the tweets. 
Taking a cue from the works related to the automatic identification of prescription drug-abuse tweets~\cite{sarker2016social,sarker2016data,klein2017detecting,phan2017enabling}, we investigated several supervised classification based approaches to filter out drug-abuse tweets from the set of tweets collected using keyword search. We proceeded with text classification as follows:

\subsubsection*{Data Annotation: }

We manually annotated $8,400$ tweets, containing $14$ different themes (see table~\ref{tab:tweet_examples}) that where identified through an exhaustive manual investigation made by $3$ annotators on $20,000$ tweets from the collected dataset. These $20,000$ were selected randomly from the collected data. We subsequently, selected $600$ unique tweets, randomly, from each theme for annotation to create a balanced annotated training set of $8,400$ tweets. Each tweet was subsequently labeled as 'Drug-Abuse (DA)' or 'Non-Abuse (NA)' by $3$ annotators and inter-annotator disagreements were solved by majority voting. The $14$ themes from table~\ref{tab:tweet_examples} can be grouped into DA or NA categories with each group having 7 themes. Hence the resulting dataset used for measuring classification performance is balanced and has equal number of drug-abuse and non-abuse tweets (i.e. $4,200$ tweets per category).

\subsubsection*{Feature Selection: }

We use three different feature sets for binary classification, based on the classifier, as follows:
\begin{enumerate}
\item \label{item:f1} A combination of bi-gram vector and handcrafted features, proposed in~\cite{klein2017detecting}.
\item \label{item:f2} A combination of semantic sentence embedding, Sent2Vec~\cite{pgj2017unsup}, and handcrafted features.
\item \label{item:f3} Word embeddings using GloVe.
\end{enumerate}

Compared to the $n$-gram based feature generation approach proposed in~\cite{klein2017detecting} that generates a large set of features (around $11,000$), the feature set generated by Sent2Vec is much smaller (around $700$).
The handcrafted features used in literature to classify drug-abuse tweets include the presence and count of $(a)$ specific abuse-indicating keywords that may indicate frequent overdoses, co-ingestion, alternative motives and routes of drug admission~\cite{hanson2013tweaking,hanson2013exploration}, and $(b)$ keywords representing drug-related slang and colloquial words\footnote{https://www.noslang.com/drugs/dictionary.php}. Each word in GloVe embedding was represented using a $100$ dimension vector.

\subsubsection*{Classification: }
\par We use 10-fold cross-validation to report classification results. In each iteration of the 10-fold, tweets in the training set and test set were sampled randomly, ensuring an equal proportion of drug-abuse and non-abuse tweets in both the sets. This was done using the “Stratified K-Folds cross-validator” library of scikit learn, which ensured that the percentage of samples from each class were preserved in training and testing set during each fold. In each fold, the sampling for the test set and training set was done with $1:9$ ratio, ensuring that the tweets from the test sets of previous iterations are not repeated in the latest test set.

We used several machine learning as well as deep neural network based methods for classification. The machine learning models included Naive Bayes, SVM, Random Forest, Logistic Regression as well as Ensemble learning techniques like AdaBoost, XGBoost, Bagging and Voting. Decision Stump was used as the base classifier for AdaBoost and Bagging, while SVM, Logistic Regression and Naive Bayes classifiers were used as the base classifiers in Voting ensemble where the majority of the label predicted by the three classifiers was considered as the label. These models were trained on two different feature sets as, mentioned in (\ref{item:f1}) and (\ref{item:f2}). The deep learning models used include LSTM, RNN,  RCNN~\cite{lai2015recurrent} as well as TextCNN~\cite{kim2014convolutional} and were trained on word embeddings mentioned in (\ref{item:f3}).

\par While LSTM and RNN are commonly used deep learning techniques for text classification, RCNN tries to improve upon them by incorporating contextual information in the recurrent structure. In contrast to the recurrent deep learning architectures, TextCNN adapts CNN for sentence classification, making it relatively faster to train and requires little hyperparameter tuning. 
All the deep learning models were trained with GloVe embeddings as inputs, where each word was represented by a 100 dimension vector. 
The LSTM model was parameterized with a dropout of $20\%$ and recurrent-dropout of $20\%$. We used bidirectional GRU as the recurrent layer in RNN model and its output was max-pooled and average-pooled. The concatenation of the max-pooled and average-pooled vectors was given as inputs to the dense layer for classification. The RCNN model was parameterized to generate $100$ dimensional (left and right) context vectors. In the TextCNN model, the kernel-size of the $3$ (parallel) convolution layers were $2$, $3$ and $5$ respectively. Each of its convolution layers had 128 filters. The output of the three convolution layers were max-pooled, concatenated and given as input to the dense layer.
Table~\ref{tab:model-params} gives a summary of the important parameters for these DL models.

\begin{table}[!hbt]
\caption{Description of the model parameters.}
\label{tab:model-params}
\centering
\begin{tabular}{|L{3cm}|C{4.5cm}|} 
    \hline 
    \textbf{Parameter} & \textbf{Value (Remarks)} \\
    \hline
    \multicolumn{2}{|c|}{\textbf{LSTM}}\\
    \hline
    Dropout & $0.2$ \\
    \hline
    Recurrent Dropout & $0.2$ \\
    \hline
    \multicolumn{2}{|c|}{\textbf{RNN}}\\
    \hline
    Recurrent Layer & Type = Bidirectional GRU (Outputs of this layer were max-pooled and average-pooled.)\\
    \hline
    Dense Layer & Input =  Concatenation of Average-pool and Max-Pool layers \\
    \hline
    \multicolumn{2}{|c|}{\textbf{RCNN}}\\
    \hline
    Left Context & dim = $100$ \\
    \hline
    Right Context & dim = $100$ \\
    \hline
    \multicolumn{2}{|c|}{\textbf{TextCNN}}\\
    \hline    
    $\#$ Parallel CNN layers & 3 \\
    \hline
    CNN Layer 1 &  Kernel size = 2 , Filters = 128\\
    \hline
    CNN Layer 2 &  Kernel size = 3 , Filters = 128\\
    \hline
    CNN Layer 3 &  Kernel size = 5 , Filters = 128\\
    \hline
    
    \hline
\end{tabular}
\end{table}

\subsubsection*{Observations: }
Table~\ref{tab:classifier-results} compares the $10$-fold cross-validation accuracy (applied on the $8,400$ annotated tweets) of different Machine Learning (ML) and Deep Learning (DL) classifiers in identifying the abuse and non-abuse tweets. 
While DL classifiers generally perform better than ML classifiers, we observe that SVM trained with a combination of sentence embeddings and hand-crafted features outperformed all the classifiers. Based on the classification performance of SVM, it was used for further classification of 2.2 million tweets. 
\par
\begin{figure}
\subfloat[True Positives \label{fig:da_correct}]{%
  \includegraphics[width=0.22\textwidth]{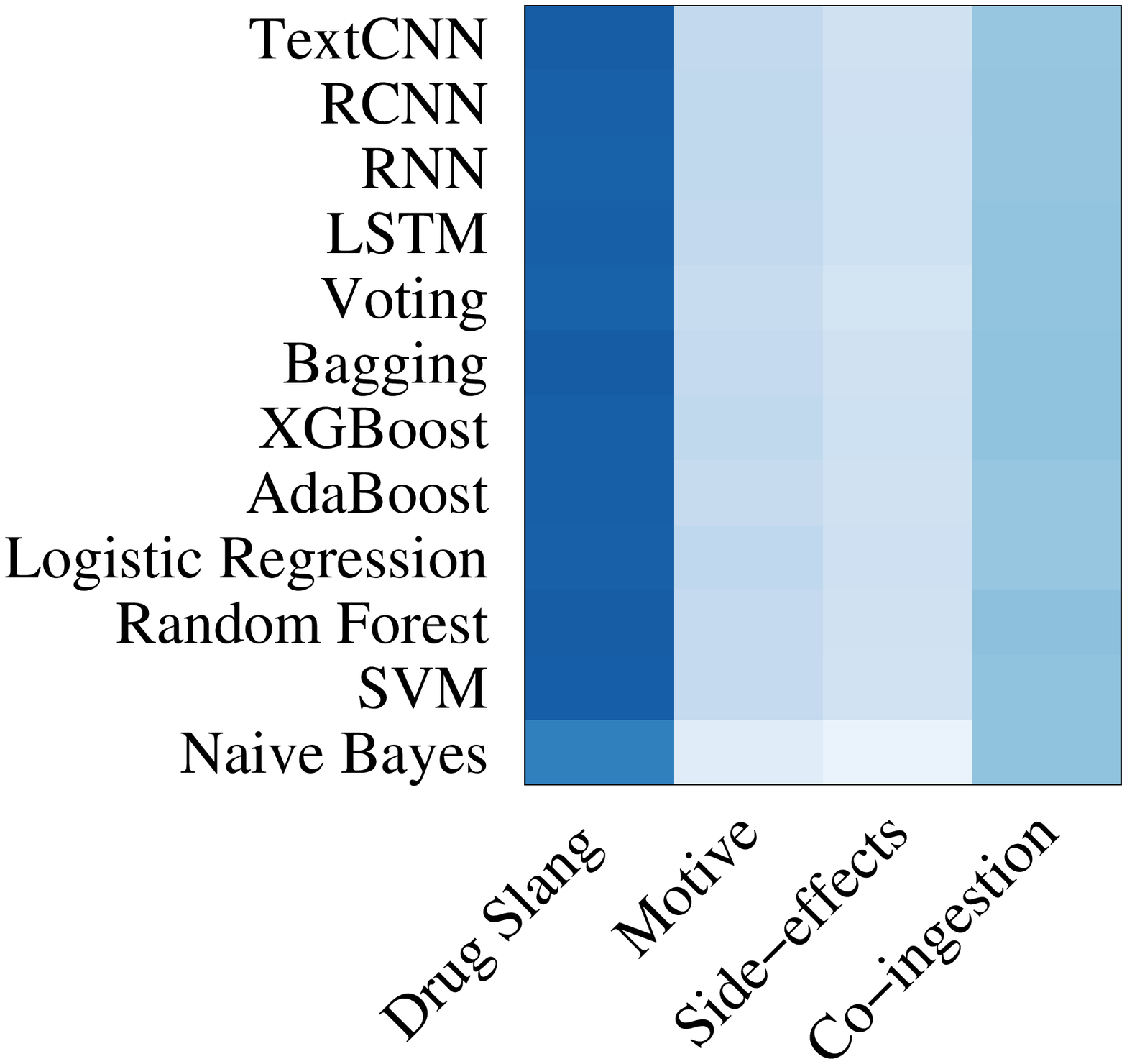}%
} 
\subfloat[False Positives \label{fig:na_incorrect}]{%
  \includegraphics[width=0.25\textwidth]{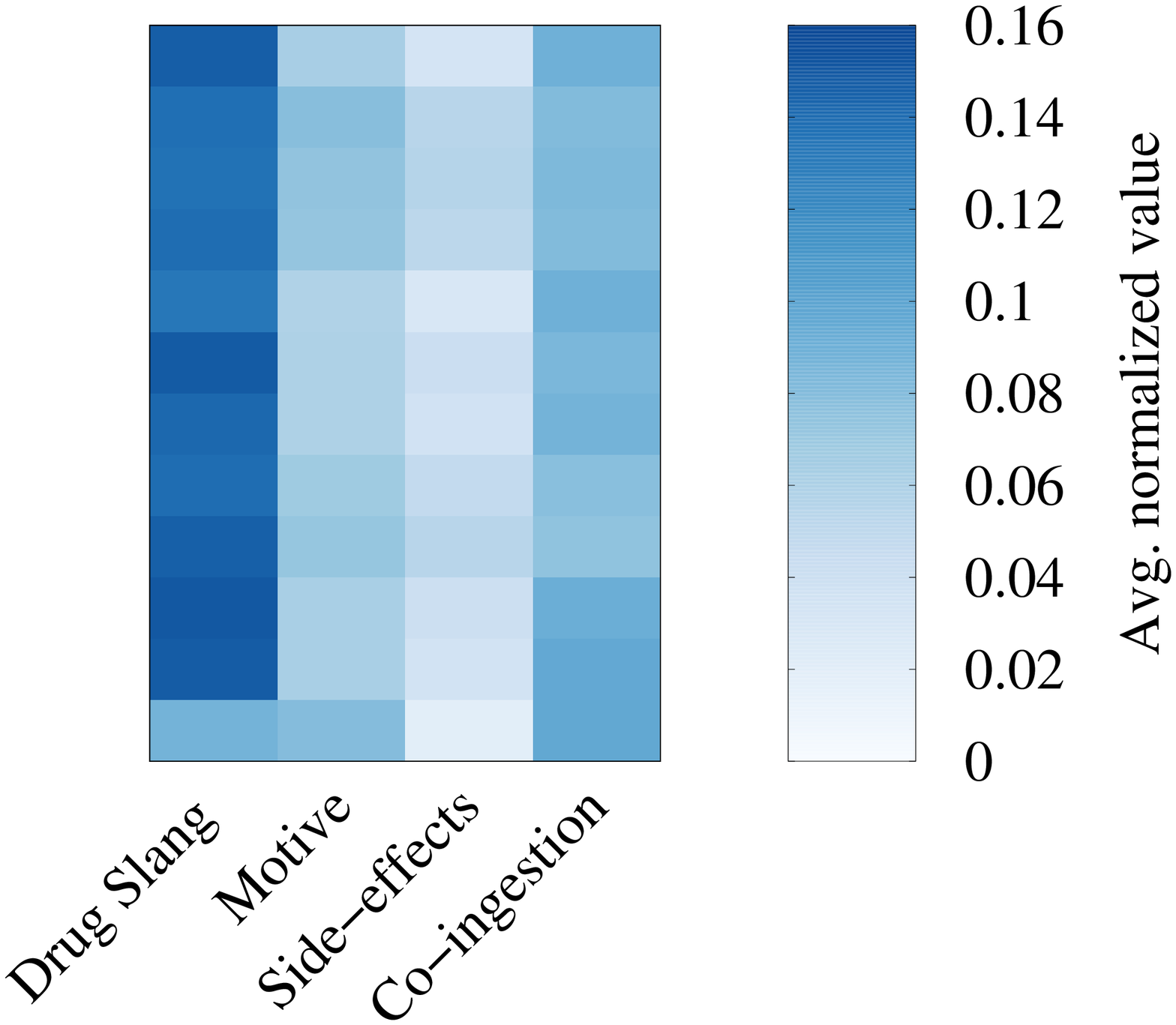}%
} \\
\subfloat[True Negatives \label{fig:na_correct}]{%
  \includegraphics[width=0.22\textwidth]{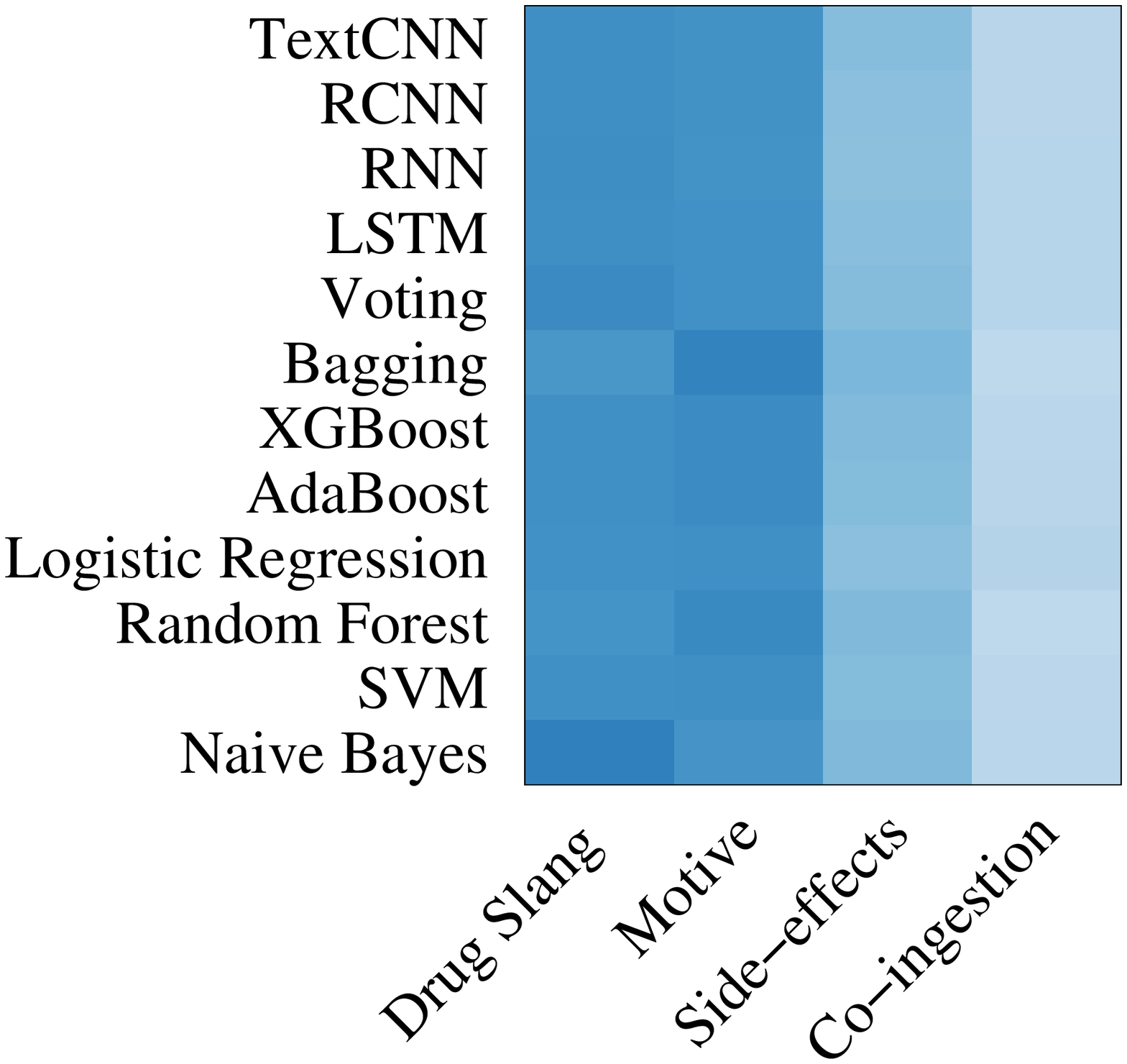}%
} 
\subfloat[False Negatives \label{fig:da_incorrect}]{%
  \includegraphics[width=0.25\textwidth]{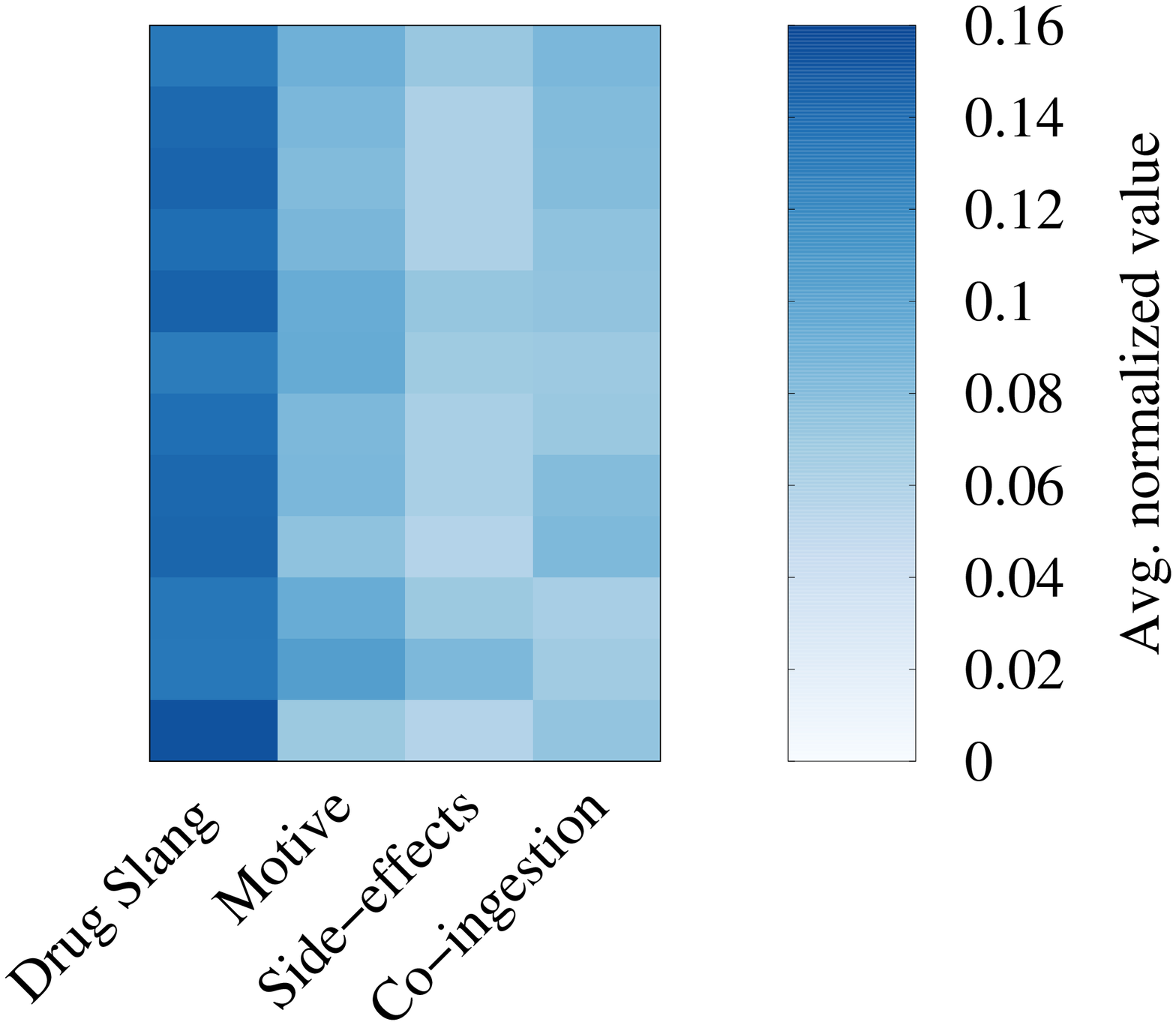}%
}

\caption{Average normalized count of prominent keywords of ~\protect\subref{fig:da_correct} correctly classified DA tweets, ~\protect\subref{fig:na_incorrect} NA tweets classified as DA, ~\protect\subref{fig:na_correct} correctly classified NA tweets and ~\protect\subref{fig:da_incorrect} DA tweets classified as NA. ML classifiers trained on sentence embeddings and DL classifiers trained on word embeddings were considered.
} 
\label{fig:misclassification}
\end{figure}

 Although the accuracy of our approach in identifying the drug-abuse tweets is significantly high, however, we investigated some of the reasons for misclassification. From figure~\ref{fig:misclassification}\subref{fig:da_correct} it is evident that the correctly classified drug-abuse (DA) tweets prominently contain drug-abuse related slang terms (like oxycotton, hillbilly, percs, etc.) and keywords hinting at co-ingestion (like alcohol, beer, etc.), while correctly classified non-abuse (NA) tweets as seen in figure~\ref{fig:misclassification}\subref{fig:na_correct} have relatively low slang terms and co-ingestion keywords but higher mentions of motive keywords (like surgery or stress) and side-effects (like insomnia or migraine). While it is difficult to determine the exact reason for misclassification, but based on the evidence it is highly likely that NA tweets might be misclassified as DA (figure~\ref{fig:misclassification}\subref{fig:na_incorrect}) due to the presence of slang terms and co-ingestion keywords.
On the other hand NA tweets might be misclassified as DA tweets due the relatively higher presence of motive terms and keywords hinting at side-effects.

\par Out of the $2.2$ million tweets, the classifier identified around $0.77$ million tweets ($36\%$ of total tweets), of $420,502$ unique users, as prescription drug-abuse tweets. We subsequently used the  Botometer~\cite{varol2017online} API~\footnote{https://botometer.iuni.iu.edu/} to identify and remove the bot accounts. This service assigns each user a bot score corresponding to the likelihood of that account being a bot. The score is calculated based on a set of $1,150$ features using account metadata, content, network and temporal information.  Using a threshold score of 0.5~\cite{varol2017online, bessi2016social}, $185$ users were labeled as bots and their corresponding tweets were discarded.

\par For ethical concerns, we follow a rigid anonymization mechanism based on guidelines mentioned in~\cite{hintze2017comparing}. The tweet and user identities were replaced by virtual identifiers. The corresponding user mentions in the tweets were also replaced by the corresponding virtual identifier. The timestamps of the tweets were also suitably replaced by relative values. All the tweets provided as examples in this paper were paraphrased and the URLs were also replaced by placeholders.
\par The observations thus highlight the enormity of the scale of the drug-abusers active in the social network and the tremendous threat they can pose in spreading of the drug-abuse menace. However, there are a few limitations of the dataset that we outline next.

\subsection{Limitations of the Dataset}
The dataset considered for this study contains information about the users, their followers and the users they follow on Twitter, in addition to each user's prescription drug-abuse tweets. The web-scraping API retrieves original tweets only, i.e., it does not contain retweets. Hence the Twitter API\footnote{https://api.twitter.com/1.1/statuses/retweets/:id.json} was used to collect retweets of drug-abuse tweets.
A limitation of this API is that it only retrieves 100 most recent retweets of the tweet. As a result, we couldn't retrieve complete retweet information of  $170$ tweets which had more than 100 retweets. Another limitation of this dataset is that it does not contain information about the time when a user followed someone else. As a result, the network created from this dataset is considered as a static network and the dynamicity of the edges could not be considered.

\subsection{Qualitative Analysis of the Tweets} 
We performed a qualitative analysis of the 2.2 million tweets that were collected in the dataset. Table \ref{tab:tweet_examples} provides representative examples of both drug-abuse and non-abuse tweets. Majority of the non-abuse tweets, (that contain drug-abuse keywords, but do not promote or report drug-abuse) can be related to spreading awareness and rehabilitation, news  or be related to reporting effectiveness or side-effects of drugs used during medical treatments. Finally, a small fraction of tweets contained references to pop culture in the form of song lyrics about drug addiction and recovery or prescription drug-abuse references in movies or television shows. 

\par As part of the study of the cascades of drug-abuse messages, we initially investigated the underlying follower network of the drug-abusers along with their engagement characteristics, both of which are major drivers of these cascades. We next describe the process of creating the follower network and describe some of its characteristics that play major roles in the spread of the contents.

\section{The Follower Network}
\label{sec:PDAN}
\par As the network provides the underlying framework for the spread of the drug-abuse tweets, we observe some of its essential properties and highlight their significance. However, we first briefly outline the steps of the formation of the network. 

\subsection{Network Formation}
\par We used the user information and their corresponding tweets (or retweets) to create the follower network of these users. For each of the $420,317$ unique users (denoted as $U$) identified from the $0.77$ million abuse tweets and retweets, we used the Twitter API\footnote{https://api.twitter.com/1.1/followers/list.json}\footnote{https://api.twitter.com/1.1/friends/list.json} to identify the \emph{follower} relation between a user and the remaining unique users. The resulting network is a directed graph represented as $G=\langle V, E\rangle$, where $V$ is the set of nodes represented by the users in $U$ and $E$ is the set of directed edges between the node pairs. A directed edge is created from node $j$ to $i$ (denoted as $e_{ij}$) if user $u_i$ is followed by $u_j$.

\par We next observe the properties of the network and investigate the possible support it can provide in spreading of drug-abuse tweets.

\subsection{Characterizing the network}

\begin{figure}
\subfloat[CCDF of followings and followers. \label{fig:follower_following-dis}]{%
  \includegraphics[width=0.22\textwidth]{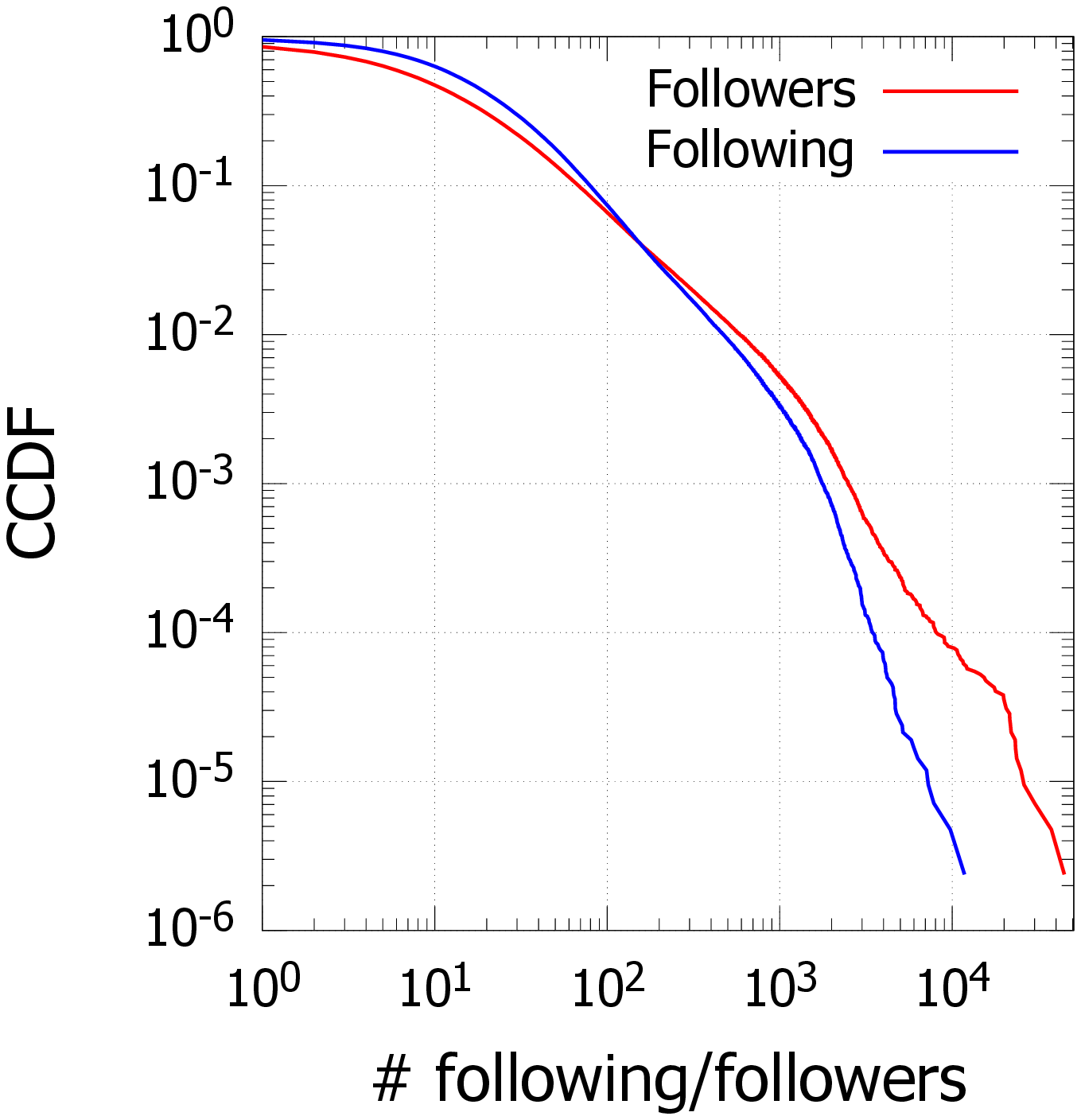}%
} \hspace{0.2in}
\subfloat[CCDF of number of tweets of users in the network. \label{fig:tweet-dis}]{%
  \includegraphics[width=0.22\textwidth]{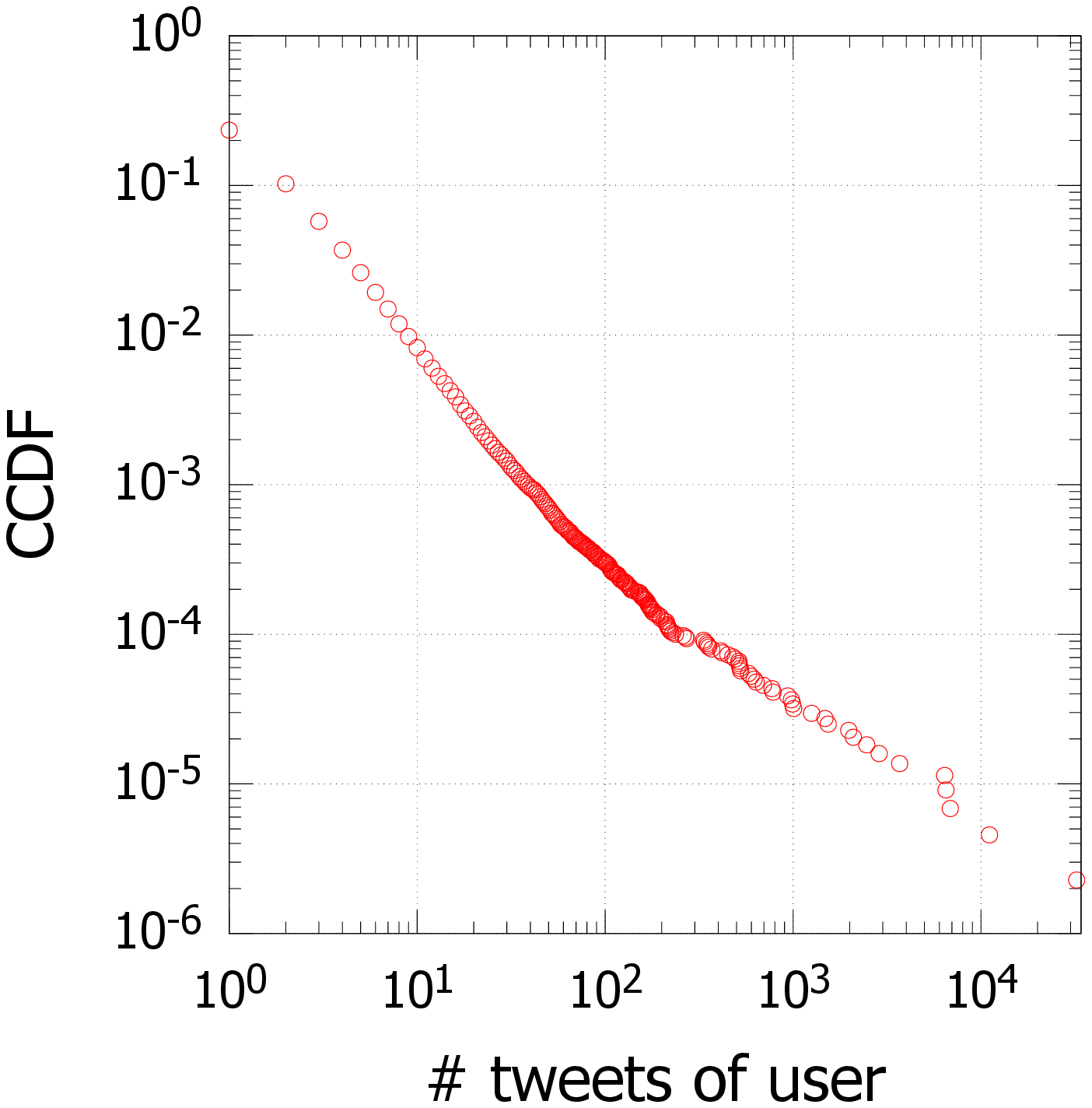}%
} 
\caption{~\protect\subref{fig:follower_following-dis} represents CCDF of the follower count and following count of the users and ~\protect\subref{fig:tweet-dis} represents CCDF of the number of tweets of users in the network. 
} 
\label{fig:network_stats}
\end{figure}

We highlight some of the significant network properties (summarized in table~\ref{tab:PDAN_Stat}) that would impact the spreading of drug-abuse tweets. 

\subsubsection*{Basic Statistics} We observed that there exists a large network consisting of approximately $0.42$ million unique users, with $17$ million links between them. The number of followers (in-degree) and followings (out-degree) of the nodes in this network follow power-law distributions with exponents $1.41$ and $1.38$, respectively. Figure~\ref{fig:network_stats}\subref{fig:follower_following-dis}, shows the complementary cumulative distribution (CCDF) of the number of  followers and followings within the network. 
The in-degree exponent indicates that a significant fraction of drug-abusers have a very high number of followers ($47\%$ of users have more than $10$  followers and $6.6\%$, i.e. $28,077$ users, have more than $100$ followers in the network), suggesting the possibility of a significant spread of the drug-abuse tweets if suitable connections exist among the nodes across the network. Consequently, we observed three major structural properties of this network --- the connectedness among the nodes, the average clustering coefficient and the reciprocity of the links --- all of which play an important role in the spread of tweets and can subsequently impact the user engagement.

\begin{table}[t]
\caption{Statistics of the follower network.}
\label{tab:PDAN_Stat}
\centering
\footnotesize
\begin{tabular}{ |l|r| } 
 \hline
  \textbf{Network Property} & \textbf{Value} \\
  \hline
  Number of nodes  & $420,317$  \\
  Number of links & $17,639,370$\\
  Average (in/out) degree & $41.97$   \\
  In-degree exponent & $1.41$   \\
  Out-degree exponent & $1.38$   \\
  Number of connected components & $237$\\
  Giant component size  & $419,700$ \\
  Clustering coefficient & $0.1516$\\
  Reciprocity & $0.6291$\\
  \hline
\end{tabular}
\end{table}

\subsubsection*{Connectedness} 
To observe the connectedness of the network, we looked at the number of connected components, i.e., subgraphs where all the nodes within it are connected through a path and have no additional connections to any other nodes outside the supergraph.
We observe that although there are $237$ connected components in the network, however the largest (giant) component comprises of $99.85\%$ of the total nodes (around $419,700$ nodes), indicating that the network is almost entirely connected. The second and the third largest connected components have $62$ and $20$ nodes respectively, and there exist several smaller components with an average size of around $2$. Further, if  we model the (giant component) network as a BowTie structure~\cite{broder2000graph} it is found that  around $91\%$ of the nodes (refer table \ref{tab:bowtie}) fall in the largest connected component (LSCC) or the core of the structure with much fewer nodes in the IN and OUT components.
A large core, apart from being resilient to targeted attacks, also implies the possibility of faster diffusion of drug-abuse tweets over a significant fraction of the network~\cite{blackJTB1966}.

\begin{table}[t]
\caption{Distribution of the users across different Bowtie components.}
\label{tab:bowtie}
\centering
\footnotesize
\begin{tabular}{|l|r|c|}
\hline
\textbf{Component} & \textbf{Count} & \textbf{Percentage}\\
\hline
LSCC& $384,027$ & $91.37$\\
IN& $29,578$ & $7.04$\\
OUT & $4,467$ & $1.06$\\
Tendrils& $1,224$ & $0.29$\\
\hline
\end{tabular}
\end{table}

\subsubsection*{Clustering Coefficient} This undirected network exhibits a high average clustering coefficient of around $0.15$ that is significantly higher than observed in the actual Twitter network ($0.096$)~\cite{AparicioEntropy15}. Although a high clustering coefficient has generally been considered as an impediment to large-scale diffusion across the networks~\cite{peresPhysica2014}, however, it needs to be investigated how the high clustering coefficient impacts the cascade properties in the drug-abuse networks.

\subsubsection*{Reciprocity} Our investigation further reveals the existence of very high reciprocity ($63\%$ of total links) in the network. This value is significantly higher than the reciprocity value of around 22\% observed in the Twitter follower network studied in ~\cite{kwak2010twitter}. Previous studies have indicated that the existence of high reciprocal links not only affects the coverage of the spread of messages in social networks but also enhances the speed of the diffusion~\cite{zhu2014influence}.

\par We later investigate the impact of these structural properties of the network on the cascades. However, these statistics point towards the existence of an underlying network platform that is amenable to the spread of the drug-abuse tweets through active engagement of a set of users. Hence we next attempt to characterize the users in  the network by their engagement pattern.  

\section{Characterizing Users in the Network}
\label{sec:user-character}
In information diffusion, users playing dominant roles in the spreading processes have often been identified based on the importance of their contents, their positional significance in the network and their engagement time. In this section, we focus on the engagement time along with the positional significance to characterize the users and deal with the contents separately in the later section. While the engagement time or activeness of the users can contribute to the speed of diffusion, the positional importance can help in increasing the breadth or depth of the cascades~\cite{qi2018diffusion}. We initially provide measures for both activeness and positional importance and subsequently characterize the users  based on both these parameters.

\subsection{Activeness of User} All users are not equally active in tweeting about prescription drug-abuse.  Figure~\ref{fig:network_stats}\subref{fig:tweet-dis} shows the complementary cumulative distribution of the number of tweets of the users.
The figure highlights that there exists a significant fraction of users with a vast number of tweets, with $4,286$ users with more than $10$ tweets and $134$ users with more than $100$ tweets. 
We consider a user as an active user if one \textit{tweets more} with low \textit{latency} (a gap between two consecutive tweets). 
The latency between two tweets is calculated based on the creation time of each tweet made available by the Twitter API. To avoid any confusion, we re-emphasize that Twitter does not provide the time when a user starts following someone.
Thus the {\em activeness score} of user $i$, denoted as $\phi_i$  is defined as follows:

\begin{equation}
\label{eq:Active}
   \phi_{i}=
\begin{cases}
    \frac{|T_{i}|}{l_{i}} ,& \text{if } |T_{i}| > 1\\ \\
    0,              & \text{otherwise}
\end{cases}
\end{equation}

where, $T_{i}=\left\lbrace t_{1}, t_{2},\dots, t_{n}  \right\rbrace$ is the set of sorted timestamps of the corresponding tweets of user $i$, $|T_{i}|$ is the total number of tweets by user $i$, and $l_{i}$ is average latency between two consecutive tweets of user $i$ that can be defined as follows:
\begin{equation}
\label{eq:avg_latency}
   l_{i}=\frac{1}{|T_{i}|-1}\left(\sum_{k=1}^{|T_{i}| - 1}(t_{k+1}-t_{k})\right)=\frac{1}{|T_{i}|-1}\left(t_{|T_i|} -t_1\right)
% \begin{cases}
%     \frac{1}{\sum_{k=1}^{|T_{i}|}\log\left(2+\left(t_{k+1}-t_{k}\right)\right)},& \text{if } |T_{i}| > 1\\
%     0,              & \text{otherwise}
% \end{cases}
\end{equation}

\begin{table}[t]
\caption{Categorizing user of the network based on their activity.}
\label{tab:Activity_Stat}
\centering
\footnotesize
\begin{tabular}{ |l|r| } 
 \hline
  \textbf{Users category} & \textbf{Number of users} \\
  \hline
  Highly active users  & $4,436$  \\
  Moderately active users & $95,281$\\
  Inactive users & $320,600$   \\
  \hline
\end{tabular}
\end{table}

\begin{figure}[!htb]
\subfloat[CCDF of user activity score in the network. \label{fig:activity_score_dis}]{%
  \includegraphics[width=0.22\textwidth]{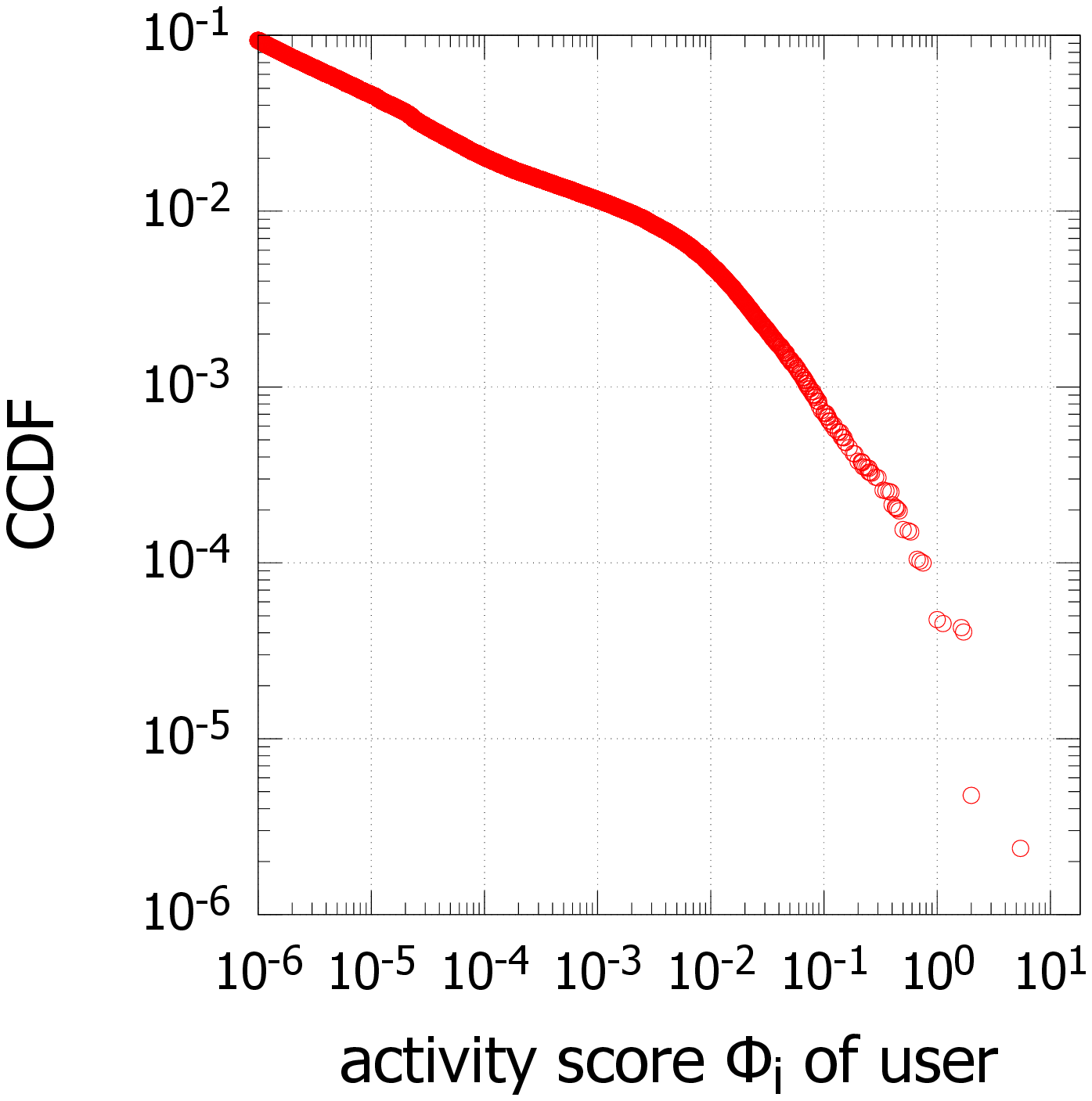}%
} \hspace{0.2in}
\subfloat[Hub score and authority score for users.\label{fig:hits_Authority_hub_score}]{%
  \includegraphics[width=0.22\textwidth]{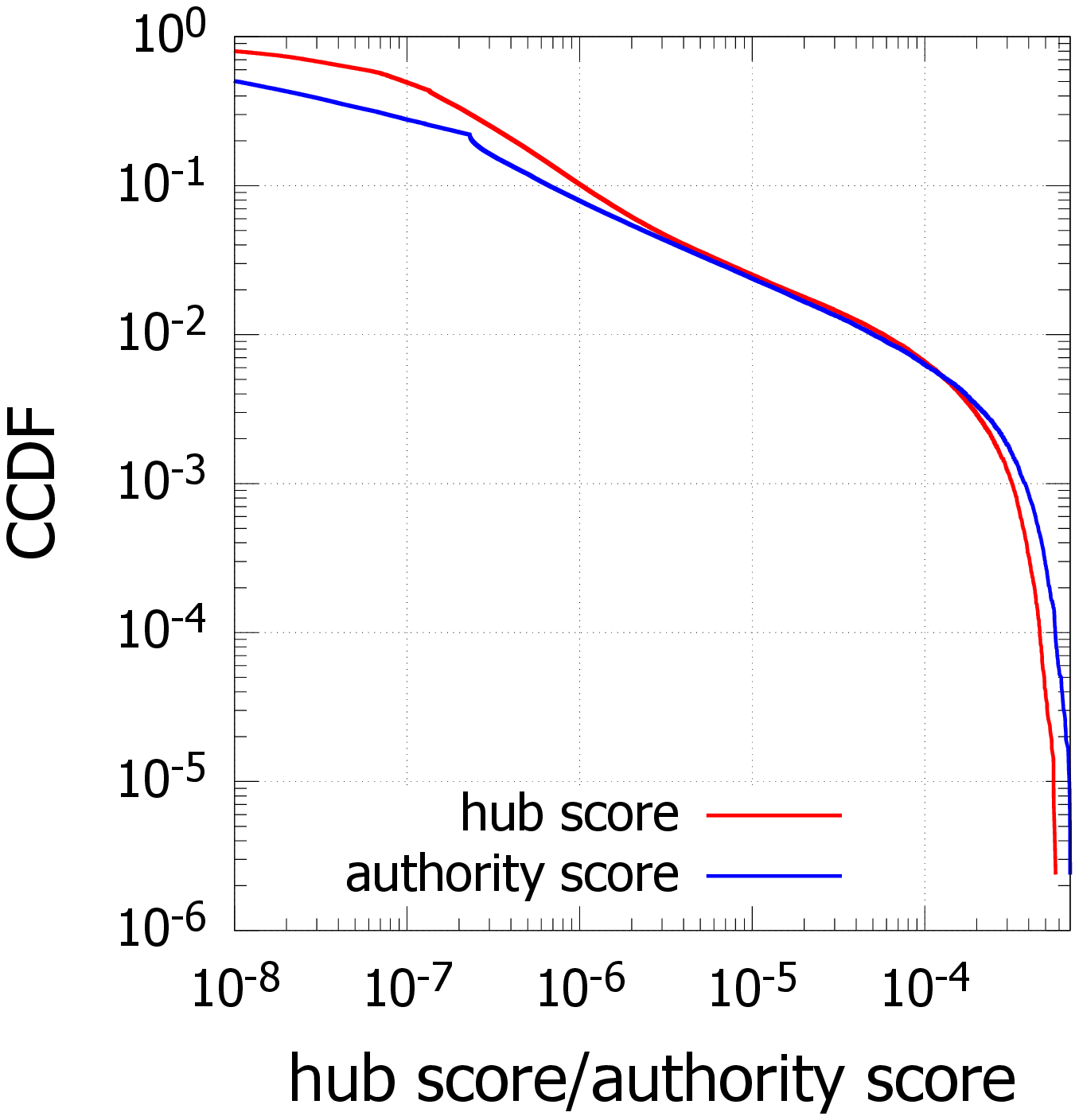}%
} 

\caption{~\protect\subref{fig:activity_score_dis} represents the CCDF of users' activity score as per equation~\ref{eq:Active}, ~\protect\subref{fig:hits_Authority_hub_score} represents CCDF of the Authority and Hub score of the users calculated using HITS algorithm.}
%\vspace{-0.2in}
\label{fig:aut_hub_dis}
\end{figure}

\begin{table*}[t]
\caption{
Reach of highly active users at each hop. Hop $0$ shows the distribution of highly active users by their role in the network followed by the distribution of users roles at each subsequent hop.
}
\label{tab:Activ_user_reach_cat}
\centering
\footnotesize
\begin{tabular}{ |r|r|r|r|r|r|c| } 
 \hline
  \textbf{Hops} & \textbf{Info.Sharing} & \textbf{Leaders} & \textbf{Info. Seeking} & \textbf{Fringe}  & \textbf{\#users reached} & \textbf{\% of network covered} \\
  \hline
  $0$ & $4$ & $88$ & $47$ & $4,297$ & $4,436$ & $1.05\%$ \\
  $1$ & $374$ & $8,828$ & $2,741$ & $79,985$  & $96,364$ & $22.93\%$ \\
  $2$ & $407$ & $3,251$ & $2,191$ & $272,548$ & $374,761$ & $89.16\%$ \\
  $3$ & $12$ & $0$ & $0$ & $37,155$  & $411,928$ & $98.00\%$ \\
  $4$ & $0$ & $0$ & $0$ & $2,256$  & $414,184$ & $98.54\%$ \\
  $5$ & $0$ & $0$ & $0$ & $130$  & $414,314$ & $98.57\%$ \\
  \hline
\end{tabular}
\end{table*}

The distribution of activity score as shown in figure~\ref{fig:aut_hub_dis}\subref{fig:activity_score_dis} is a heavy-tailed power-law distribution. To categorize the users based on the activity score, we used the Head/Tail breaks algorithm \cite{jiang2013head} to cluster the distribution into 2 parts. Users with $\phi_i>1.4\times10^{-3}$ (i.e. the tail) were classified as \textit{highly active} users. The remaining users were further classified into 2 categories, \textit{moderately active} ($0<\phi_i\leq1.4\times10^{-3}$) and \textit{inactive} ($\phi_i=0$). Table~\ref{tab:Activity_Stat} shows the number of users belonging to each category according to activity score.

\subsection{Positional Importance} 
The position of a user in the network can determine her reachability (ability to reach a broad set of users through her tweets) as well as her accessibility (ability to receive tweets from a large number of users). To capture both these characteristics simultaneously, we calculated the hub and authority score of each user in the network. Authority score of a node is high if it is followed by nodes with high hub score, whereas the hub score of a node would be high if it follows nodes with high authority scores. 
Thus, while authorities can act as good information spreaders because of their followers, hubs can act as information collectors obtaining diverse information from different authorities. We use the HITS algorithm~\cite{kleinberg1999authoritative} to calculate authority and hub score of each user.

\begin{table}[t]
\caption{Properties of different user roles.}
\label{tab:role_properties}
\centering
\footnotesize
\begin{tabular}{|l|r|c|c|}
\hline
\multirow{2}{*}{\textbf{User Role}} & \multirow{2}{*}{\textbf{\#Users}} & \textbf{Mean} & \textbf{Mean} \\
& & \textbf{in degree} & \textbf{out degree} \\
\hline
Fringe & $402,348$ & $24.97$ & $27.87$\\
Info. seeking & $4,979$ & $72.48$ & $133.06$ \\
Leaders & $12,167$ & $467.51$ & $466.09$\\
Info. sharing & $823$ & $1877.20$ & $ 110.32$\\
\hline
\end{tabular}
\end{table}

Figure \ref{fig:aut_hub_dis}\subref{fig:hits_Authority_hub_score} shows the CCDF of the hub and authority scores. We labeled the users as high authority user if its authority score is above $6\times10^{-6}$ (obtained using Head/tail breaks algorithm \cite{jiang2013head}) and the rest of the users are labeled as low authority users. Those users having hub score above $4\times10^{-6}$ using the same algorithm are labeled as high hub user, and the rest of the users are labeled as low hub users.

\par We categorized the users into four role types based on authority and hub scores:~\emph{\textbf{a) information seeking}} -- who have high hub and low authority scores,~\emph{\textbf{b) information sharing}} -- who have high authority and low hub scores,~\emph{\textbf{c) leaders}} -- who have high hub as well as high authority scores and ~\emph{\textbf{d) fringe}} -- who have low hub and low authority scores.

In table~\ref{tab:role_properties}, we observed that around $95\%$ of the total users are fringe nodes who have few followers as well as very few followings. 
On the other hand, the total number of users in each of the remaining three categories is only $1-2\%$. The users in the three remaining categories represent the influential section of users in the network who have the capability to spread drug-abuse tweets across a large section of the network. Thus it is necessary to investigate the contribution made by each of these user types in the spread of drug-abuse tweets; hence we next correlate the activity score of the users with their hub and authority scores to explore their potential in spreading drug-abuse tweets.

\subsection{Characterizing Highly Active Users} Initially, we took a  closer look at the active users and observed their role types. The first row in table~\ref{tab:Activ_user_reach_cat} shows the number of active users in each of the role categories. As can be observed, more than $96\%$ of the highly active users are fringe nodes. This indicates that even though fringe nodes hold less positional importance but a significant fraction of the active engagements in generating drug-abuse contents is made by them.
Thus to investigate the fraction of users getting exposed to the drug-abuse tweets generated by these active users, we further observed the reach of the active nodes at different hops.

\par As shown in table~\ref{tab:Activ_user_reach_cat}, the active nodes by themselves can reach only $23\%$ of the network in the first-hop but manage to reach $89\%$ of the network in the second-hop. Thus, the reachability of the first-hop neighbors provides the active users the potential to reach the bulk of the nodes in the network. On taking a look at the properties of the first-hop neighbors, we find that around $13\%$ of these nodes have either a high hub or authority score i.e. they are non-fringe users, including more than $9\%$ leaders who have both high hub and authority scores. Roughly $0.21$ million or $77\%$ of the users in the second-hop follow at least one non-fringe user in the first-hop, highlighting the importance of the connectivity of non-fringe nodes.
Interestingly, we also observed that $81\%$ of the first-hop neighbors follow one or more active fringe nodes, indicating that even though individually the active fringe nodes are not structurally important but collectively they can reach a significant number of users in the first-hop.

\par Thus a major takeaway from these observations is that although most of the active users are positionally fringe, however, due to the positional importance of their first-hop neighbors, drug-abuse discussions initiated by these nodes can potentially reach a significant population of users in the network. We next observe the actual cascades and investigate the role of the underlying network along with the key players and the tweet contents in the spreading process.
\section{Cascades and Spread}
\label{sec:spread}

\par To measure the extent of the spread and influence of tweets, we investigate the cascades formed through drug-abuse discussions. We discovered the key players along with the pattern of user engagement responsible for the formation of  the cascades. Here user engagement refers to both generation of new drug-abuse tweets as well as re-tweets by the users. Subsequently, the dynamics of spread with respect to the drug names are investigated, keeping in mind the complex contagion phenomenon that is typical to the spread of social behavior.

\subsection{Measuring Cascades}

We initially describe the experimental procedure to identify the cascades followed by the measures of various properties of the cascades.
\subsubsection*{Dataset Preparation} 
The cascades might be formed when users directly retweet a drug-abuse tweet or create a fresh content based on one that has appeared in their timeline. While retweets can directly be linked to a cascade, determining whether a new drug-abuse tweet has been made based on the previously received tweet (and hence should be a part of the cascade) is difficult. Possibilities exist that the user might have been influenced by certain external sources and not the drug-abuse tweet she has received before tweeting. However, ignoring such tweets entirely as not being part of the cascades can lead to a severe undermining of the veracity of the cascade problem. Hence to reduce the chances of such \textit{coincidental errors}, following assertions were made: a) a tweet $T$ was added to the cascade if at least one previous tweet from the cascade appeared on the user's timeline within a short period, $\Delta$, prior to the sending of $T$ and b) we consider only large cascades  for analysis so as to ensure that a significant fraction of the spread does not suffer from such error. 
The value of $\Delta$ was chosen as nine days based on the study in~\cite{icwsmpersistencekleignberg}, where it is shown that the mean of the attention decay time (time between peak attention and $75\%$ of attention) of the tweets is around $217$ hours.
The process of identifying cascades is formally defined below.

\par The drug-abuse tweets in the dataset are initially sorted based on their time of generation. For each tweet, the corresponding creator is identified and is included as the initial node in the cascade graph $G=\langle V, E\rangle$. A user, $v$, following the initiator $i$ is added as a node to the cascade graph if it has either re-tweeted or created a new drug-abuse tweet within nine days of the appearance of the parent tweet in her timeline.  A directed link is created from node $i$ to the follower $v$, indicating that engagement of node $v$ has possibly been influenced by $i$. This process is further recursively repeated for the followers of the newly added users in $G$. By following this process, we extracted around $50,409$ cascades of different lengths. 
We focus on the large cascades as they mainly reflect the threat posed by social media in the spread of drug-abuse tweets. To study the characteristics of the large cascades, we considered cascades of length $\geq 20$ for our investigations. The choice of this value is not a principled one but is based on our observation that users with different characteristics play a consistent role across cascades beyond the length of $20$.
We next investigate the key structural properties of these large cascades.

\subsubsection*{Structural Properties of the Cascades} 

\begin{figure}[t]
\subfloat[CCDF of cascade size. \label{fig:ccdf_cascade_size}]{%
  \includegraphics[width=0.18\textwidth]{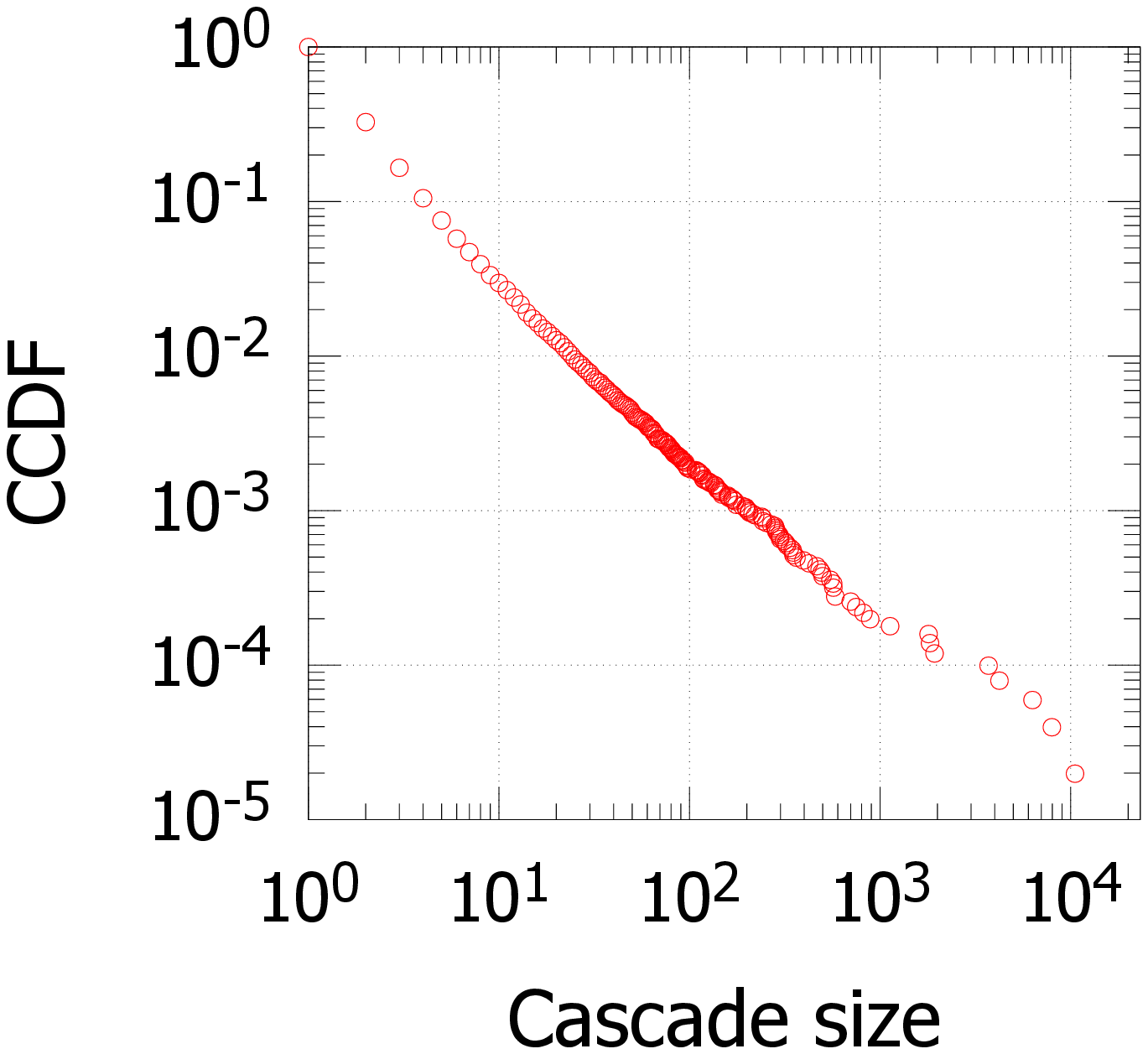}%
} 
\subfloat[Structural Virality. \label{fig:structural_virality}]{%
  \includegraphics[width=0.30\textwidth]{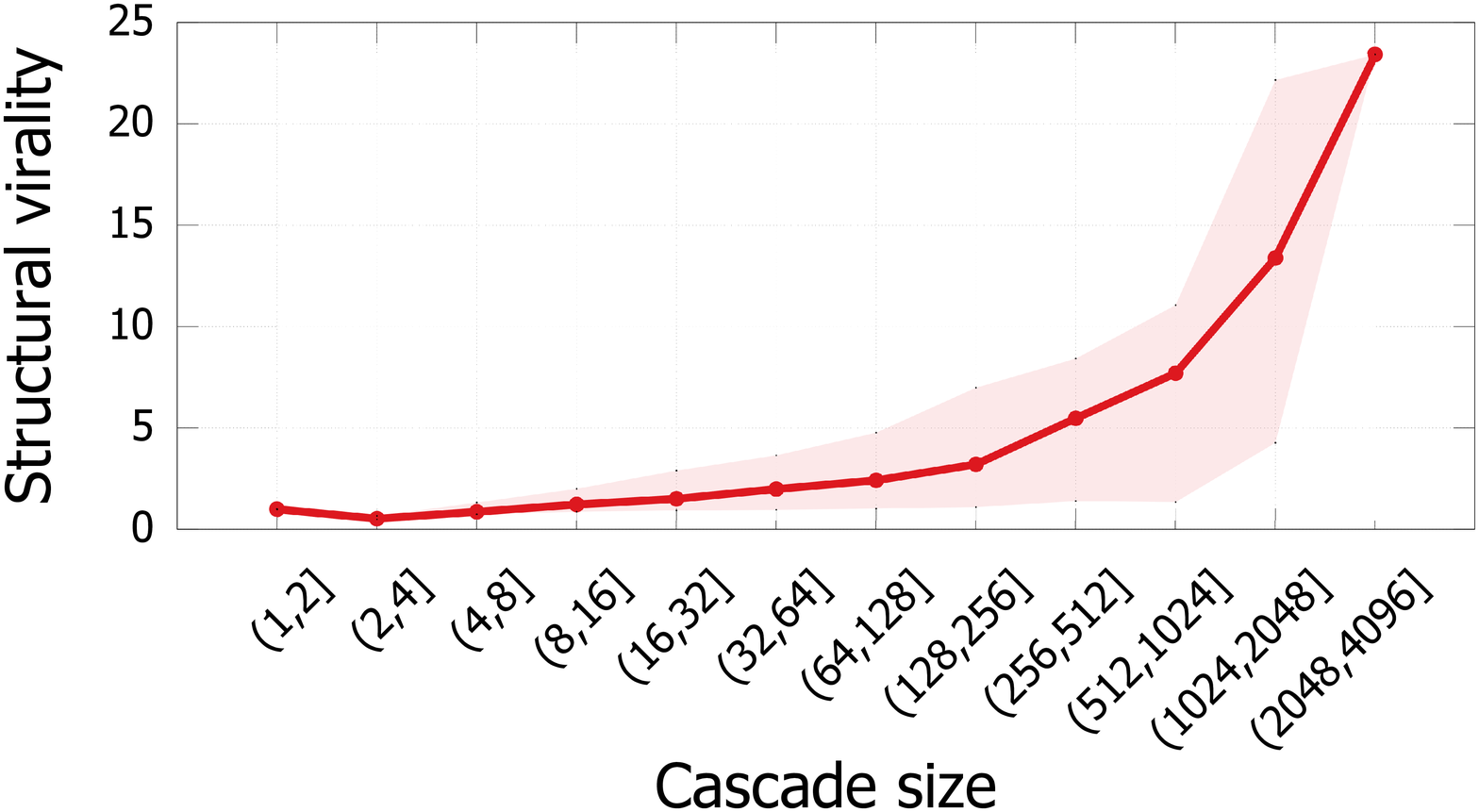}%
}
\caption{~\protect\subref{fig:ccdf_cascade_size} represents CCDF of the cascade size ~\protect\subref{fig:activity_score_dis} represents the relation between cascade size and mean structural virality (Wiener index) for each bin.}
%\vspace{-0.2in}
\label{fig:virality}
\end{figure}

The distribution of the cascade sizes provides an idea about the scale of user engagement. As shown in figure~\ref{fig:virality}\subref{fig:ccdf_cascade_size}, the distribution of the cascade sizes follows a power-law with an exponent of $2.06$. The maximum cascade size observed is $23,164$, which  indicates large cascades of user engagement may be formed due to the spread of drug-abuse tweets.

\par We also observed the structural virality of these cascades, as defined in~\cite{andersonManSc2015}. 
The structural virality has been measured using the Wiener index that is given by the average shortest path length ($d_{avg}$) between any pair of nodes in the cascade graph. A lower value of $d_{avg}$ (near to $2$) indicates a hub-like structure where a single powerful node causes the entire cascade, whereas larger values indicate viral diffusion through branches involving multiple propagating nodes. Figure~\ref{fig:virality}\subref{fig:structural_virality} shows the distribution of the structural virality observed in the follower network. As can be observed, the structural virality steadily increases with increasing cascade size. This indicates that multiple nodes play an important role in the cascades.

\subsubsection{Subgraph Properties of the Cascade Nodes}

\begin{figure}[t]
\subfloat[Avg. Path Length and Diameter. \label{fig:cascade_path_properties}]{%
  \includegraphics[width=0.22\textwidth]{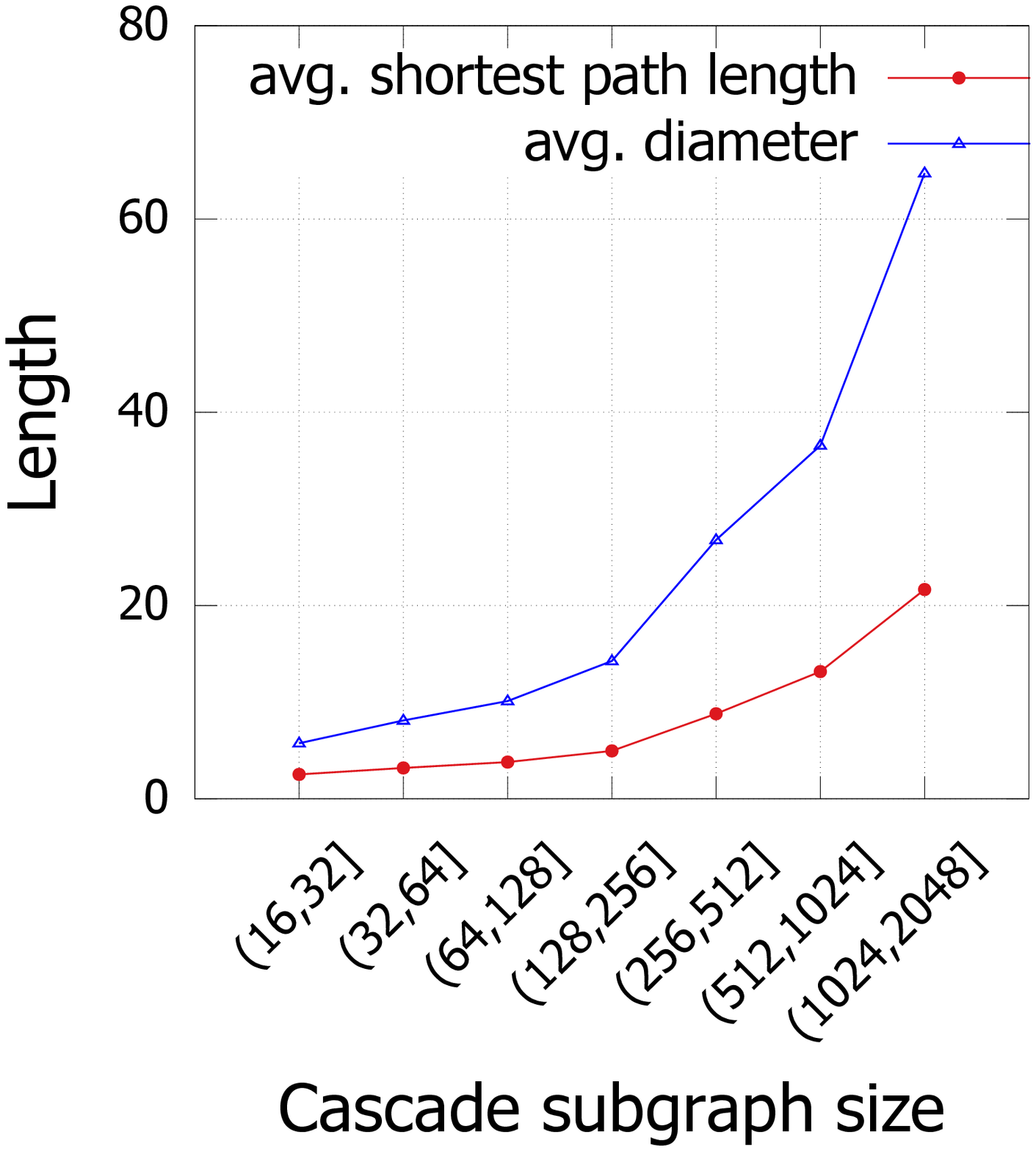}%
} \hspace{0.2in}
\subfloat[Avg. Clustering Coeff. and Reciprocity. \label{fig:cascade_edge_properties}]{%
  \includegraphics[width=0.22\textwidth]{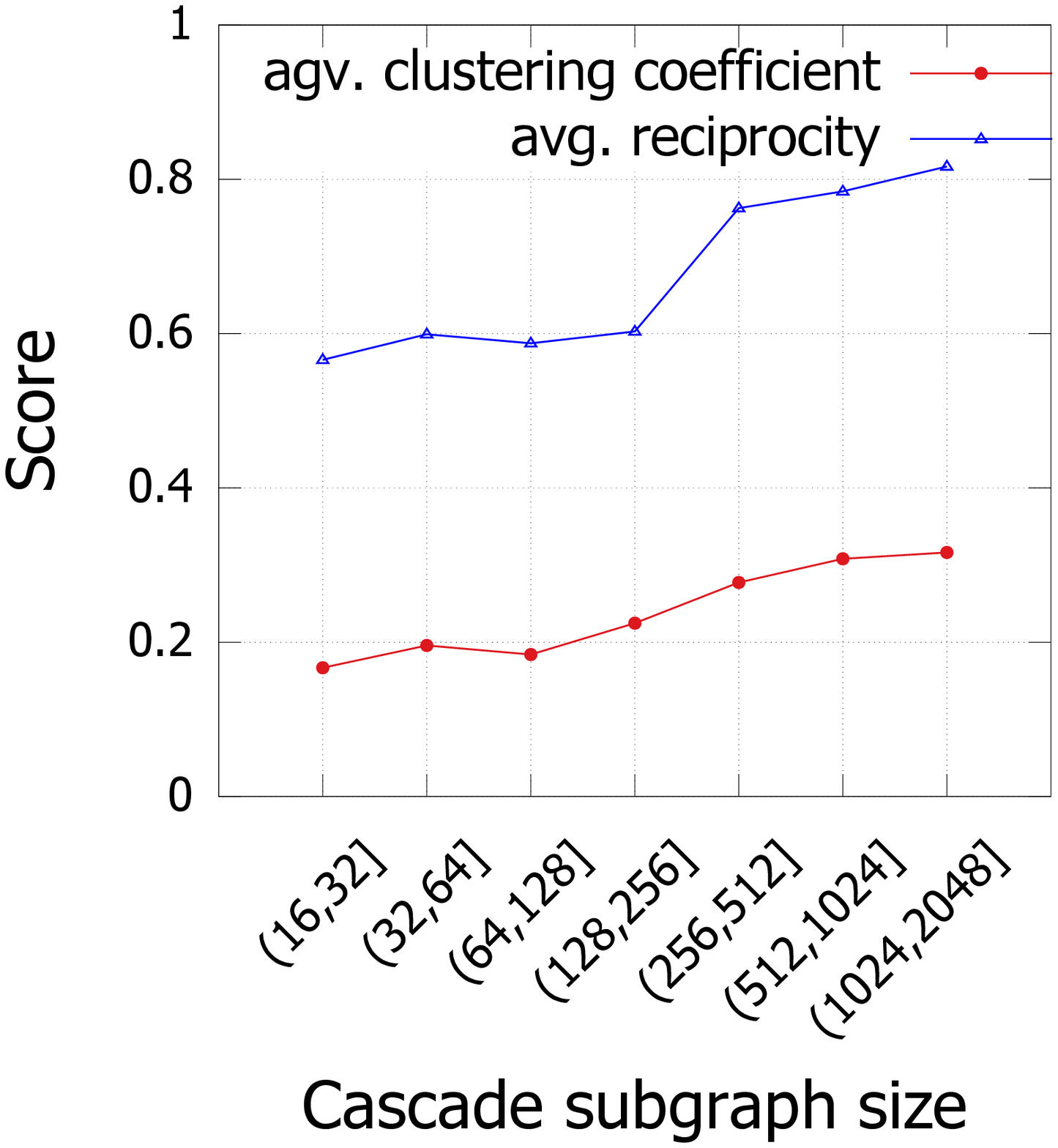}%
}
\caption{Properties of cascade nodes in network subgraph. ~\protect\subref{fig:cascade_path_properties} shows the average shortest path and diameter of the cascade nodes in network subgraph and, ~\protect\subref{fig:cascade_edge_properties} shows the average reciprocity and clustering coefficient.}
%\vspace{-0.2in}
\label{fig:cascade_properties}
\end{figure}

To study the network structure of the nodes involved in the large cascades, for each cascade we constructed a network subgraph that comprises of the nodes of the corresponding cascade. As shown in figure~\ref{fig:cascade_properties}\subref{fig:cascade_path_properties}, the average path length between the nodes in the subgraph vary from $2.51$ to $21.66$, whereas the corresponding average diameter ranges from $5.74$ to $64.75$ with respect to the subgraph sizes. These values indicate that these cascades reach far beyond the immediate neighbor nodes. It is also observed that the mean reciprocity and clustering coefficient of these subgraphs are significantly high (figure~\ref{fig:cascade_properties}\subref{fig:cascade_edge_properties}). The mean reciprocity values with respect to the subgraph sizes vary between $0.57$ and $0.81$, whereas the mean clustering coefficients range between $0.17$ and $0.31$.  A high average clustering coefficient and reciprocity of the cascade nodes provide strong evidence that the cascades spread  along the follower chain through groups of closely connected nodes. 

\par We next focus on the users to identify the key players involved in the spread of the drug-abuse tweets.

\begin{figure}[ht]
\centering
\subfloat[CCDF of node properties.\label{fig:high_activity}]{%
 \includegraphics[width=0.22\textwidth]{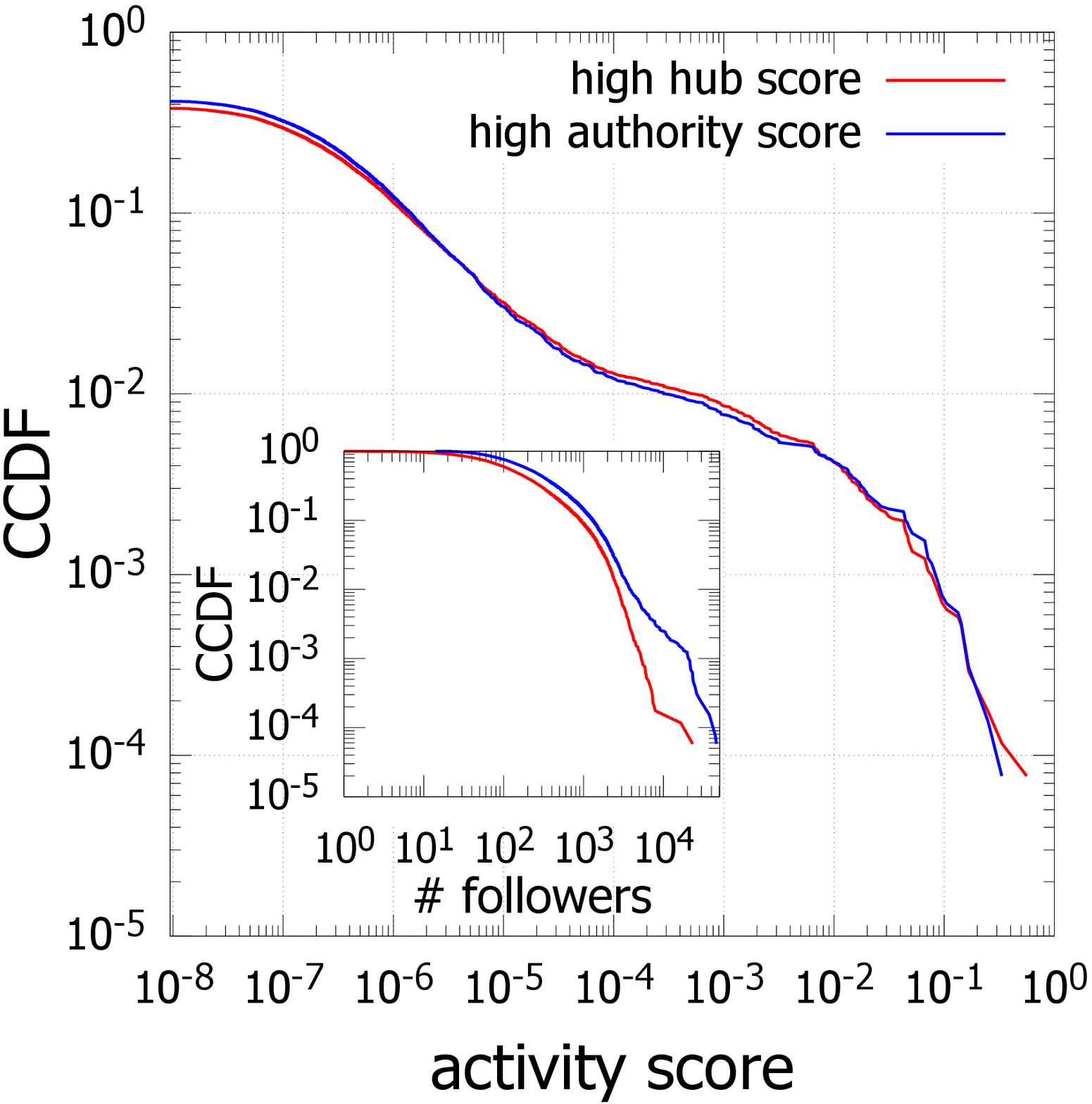}%
} \hspace{0.2in}
\subfloat[User distribution in the network. \label{fig:information_paradox}]{%
  \includegraphics[width=0.22\textwidth]{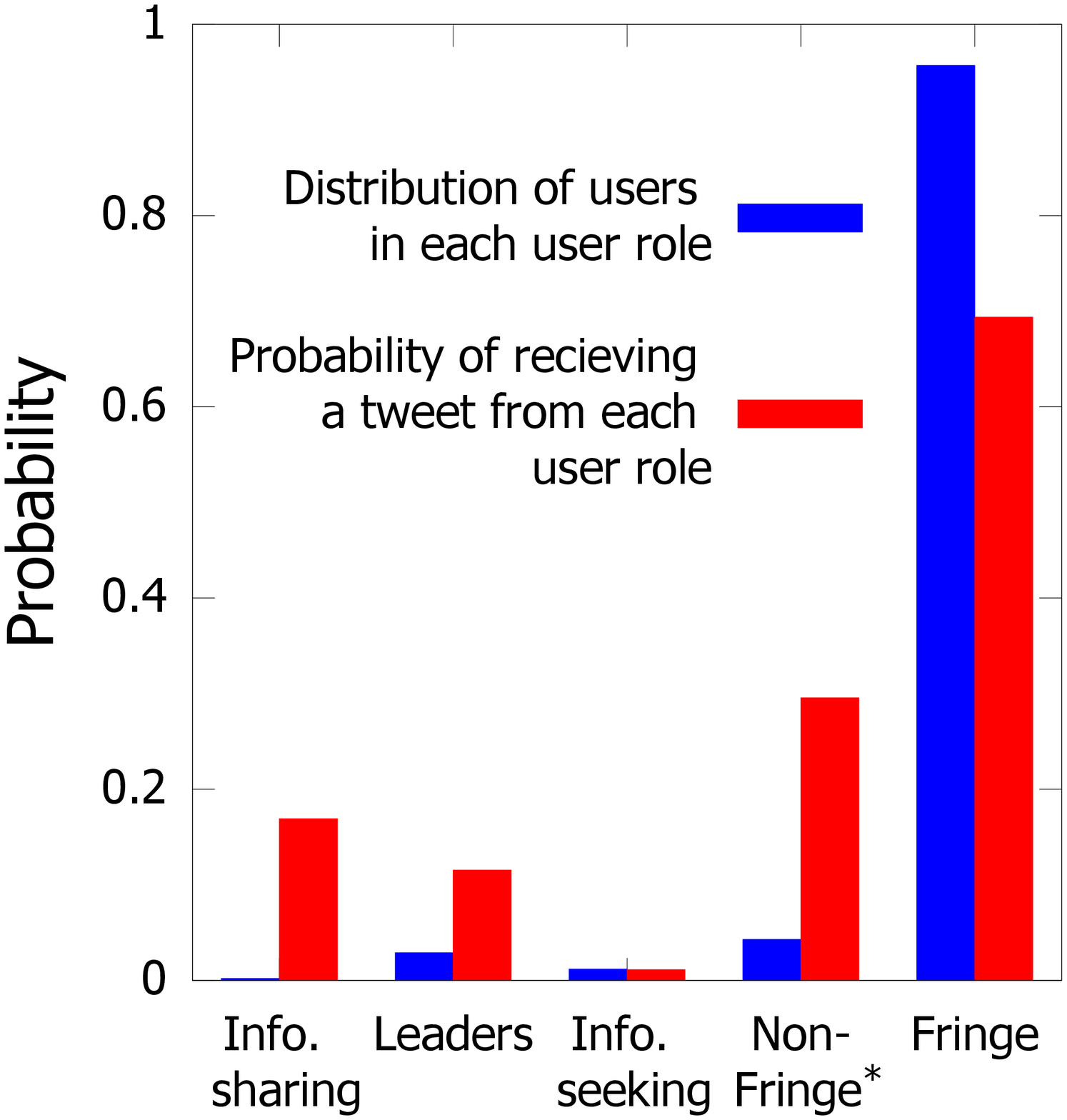}%
}
\caption{~\protect\subref{fig:high_activity} represents CCDF of activity score and the inset in ~\protect\subref{fig:high_activity} represents CCDF of the \#followers of the high hub and authority score users. ~\protect\subref{fig:information_paradox} gives the probability distribution of users in each category and the probability distribution of receiving a drug-abuse tweet from each user category. Non-Fringe category represents the combined probabilities of info. sharing, leaders and info. seeking users. 
}
\label{fig:high_hits_users}  
\end{figure}

\begin{table*}[ht]
\caption{Comparison of the properties of cascades initiated by fringe and non-fringe nodes. The category of users involved play an important role in determining the properties of the cascade they belong to.}
    \label{tab:user_role_cascades}
\centering
\scriptsize
\begin{tabular}{| l | C{1.28cm} | C{1.28cm} | C{1.28cm} || C{1.5cm} | C{1.5cm} | C{1.5cm} |}
		\hline
    	\multirow{2}{*}{\textbf{Cascade Property}} & \multicolumn{3}{c||}{\textbf{Cascades initiated by fringe nodes}} & \multicolumn{3}{c|}{\textbf{Cascades initiated by non-fringe nodes}} \\
        \cline{2-7}
         & Value & $\sigma$ & Percent & Value & $\sigma$ & Percent \\
        \hline
        \#cascades initiated  & $487$ & -  & $71.72\%$ & $192$ & - & $28.28\%$ \\
        Avg. cascade size & $177.78$ & $1237.27$ & - & $106.67$ & $458.51$ & - \\
        Avg. structural virality ($d$) &  $2.63$ & $3.95$ & - &  $1.97$ & $1.44$ & - \\ 
        Max depth of cascade & $313$ & - & - & $84$ & - & - \\
        Avg. depth of cascades & $10.72$ & $22.59$ & - & $6.83$ & $8.41$ & - \\
        \multicolumn{1}{|m{3.5cm}|}{Avg. depth at which max. width was observed  in cascade} & $5.95$ & $14.84$ & - & $3.02$ & $4.61$ & - \\
        \hline
        %Avg. First hop nodes in cascade & $1,269$ & - & $2,057$ & - \\
        Avg. \#first hop nodes in cascade & $5.05$ & $8.75$ & - & $28.05$ & $37.29$ & - \\
        Avg. \#first hop non-fringe nodes & $0.27$ & $0.53$ & $5.36\%$  & $1.44$ & $1.59$ & $5.14\%$ \\
        Avg. \#first hop fringe nodes & $4.77$ & $8.79$ & $94.64\%$ & $26.60$ & $37.11$ & $94.86\%$ \\
        \hline
        Avg. \#second hop nodes & $8.14$ & $22.68$ & -  & $7.87$ & $11.08$ & - \\
        \multicolumn{1}{|m{3.5cm}|}{Avg. \#second hop nodes with non-fringe node as parent} & $3.64$ & $22.03$ & $44.77\%$ & $3.39$ & $ 9.25$ & $43.07\%$ \\
        \multicolumn{1}{|m{3.5cm}|}{Avg. \#second hop nodes with fringe node as parent} & $4.49$ & $7.52$ & $55.23\%$  & $4.48$ & $6.70$ & $56.93\%$ \\
        \hline
	\end{tabular}
\quad
\end{table*}

\subsection{Key Players in Spreading}
\begin{comment}
\begin{figure}
\centering
\includegraphics[width=0.4\textwidth]{images/cascade_size_vs_inactive_grouped.pdf}
\caption{Average percentage of inactive users in cascades. Cascades more than 20 nodes are considered.}
\label{fig:inactive-percent}
\end{figure}
\end{comment}
We investigated the key players in the cascades considering both the activeness as well as the positional importance of the users. Since a large fraction of the users in the network is largely inactive, we initially observed the activeness of the users in the cascade. We observed that for cascades involving more than $20$ users, around $50-60\%$ of the users are {\em one time engagers} who had contributed only one drug-abuse tweet but helped in keeping the cascade alive. This  reveals the importance of these inactive users who  contribute to increasing the length of the cascades. 

\par We next observe the role of the fringe and non-fringe nodes in the spreading process. 
Although it has been observed that the non-fringe active nodes are significantly less in number (refer table~\ref{tab:Activ_user_reach_cat}) as compared to the fringe ones,
however, it is observed that around $30\%$ of the drug-abuse tweets (as seen in figure~\ref{fig:high_hits_users}\subref{fig:information_paradox}) 
in the timeline of a random user are generated by a non-fringe node, even though they constitute only $5\%$ of the nodes in the network. 
A closer look at table~\ref{tab:role_properties} %and \ref{tab:role_out_degree_distr} 
reveals that in-degree of non-fringe nodes are $17$ times higher (computed using weighted average) compared to in-degree of fringe nodes. So the probability that a node follows a non-fringe node is comparable with the probability that the node follows a fringe node. 
Hence, as the fraction of  fringe and non-fringe nodes followed by a random user is comparable, the contents generated by both these node types in the timeline of a random user is also comparable. This indicates that both the non-fringe as well as the fringe nodes are similarly responsible in the spread of these drug-abuse tweets.

\begin{figure*}[!htb]
\subfloat[Overall\label{fig:persistance_overall}]{%
  \includegraphics[width=0.2\textwidth]{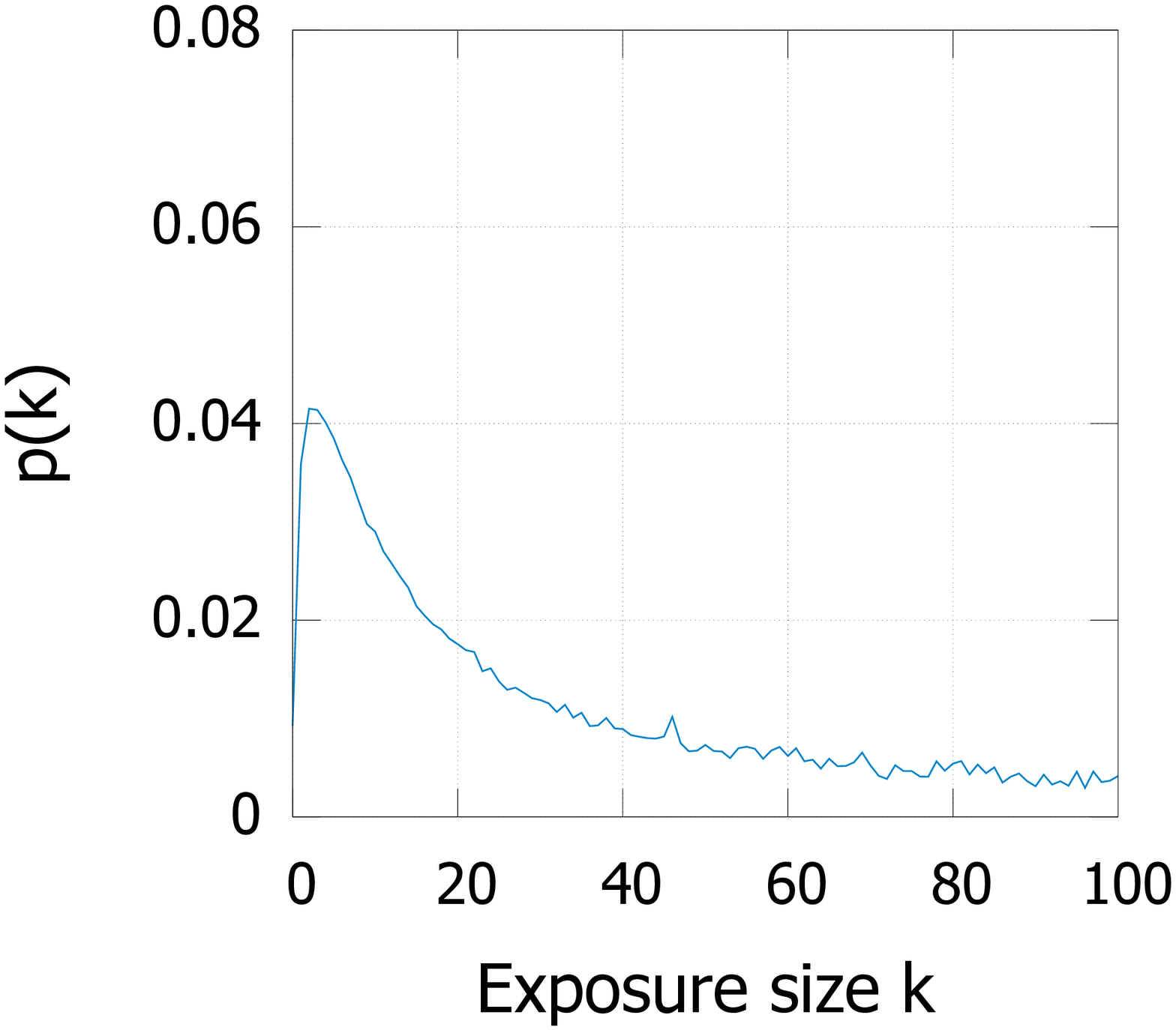}%
} 
%\hspace{0.1in}
\subfloat[Vicodin\label{fig:persistance_vicodin}]{%
  \includegraphics[width=0.2\textwidth]{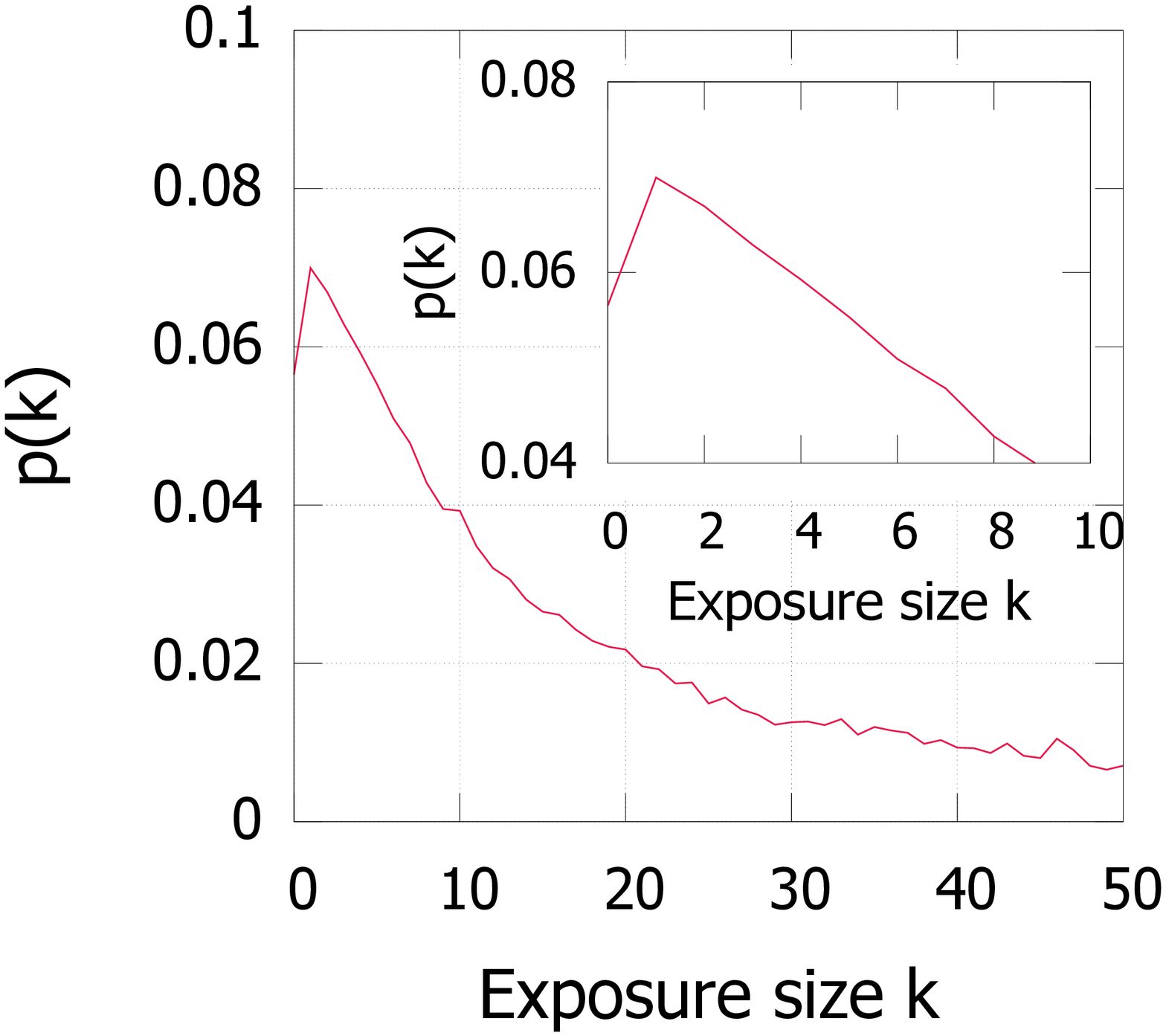}%
} 
%\hspace{0.1in}
\subfloat[Percocet \label{fig:persistance_percocet}]{%
  \includegraphics[width=0.2\textwidth]{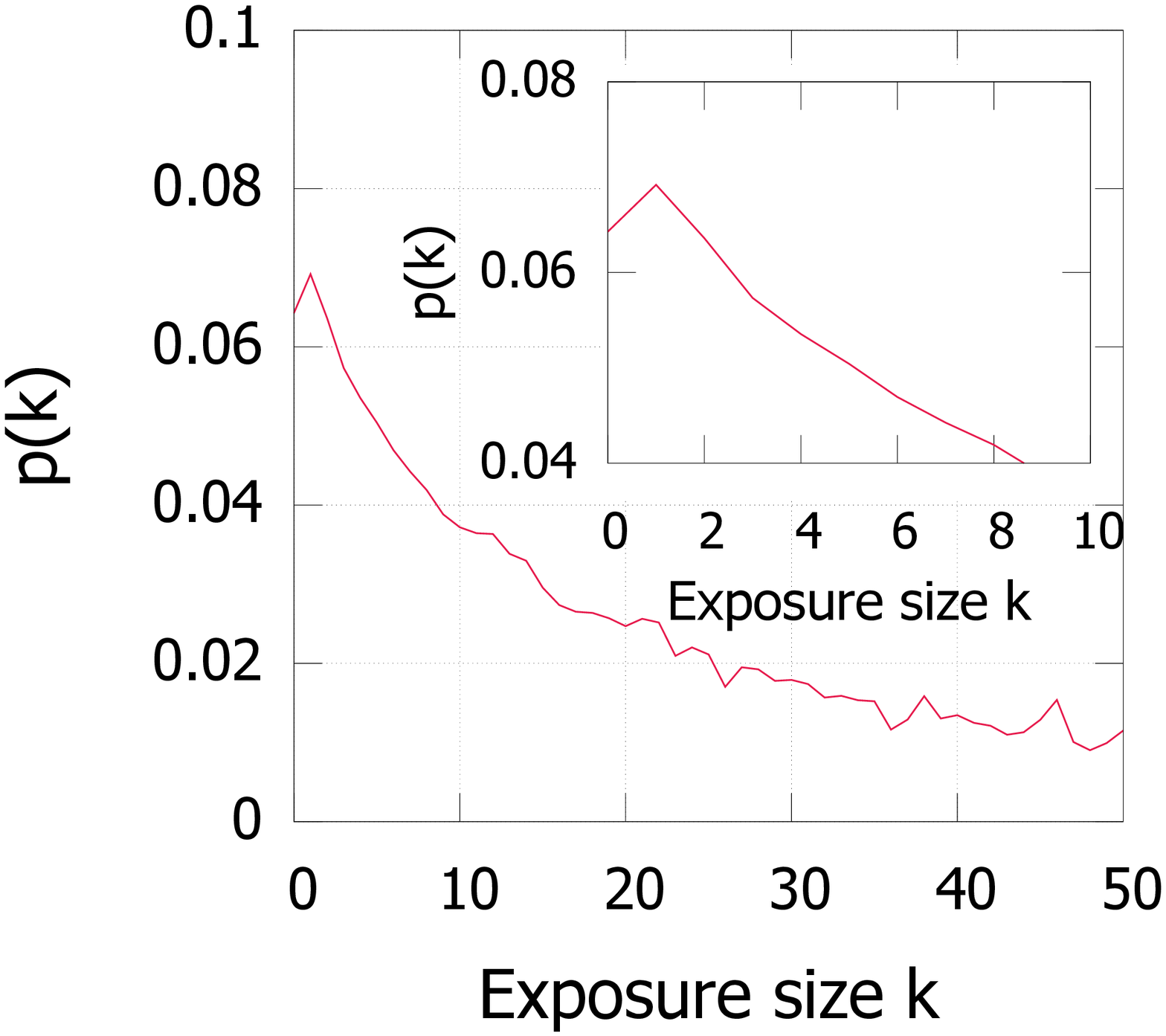}%
} 
%\hspace{0.1in}
\subfloat[OxyContin \label{fig:persistance_oxycontin}]{%
  \includegraphics[width=0.2\textwidth]{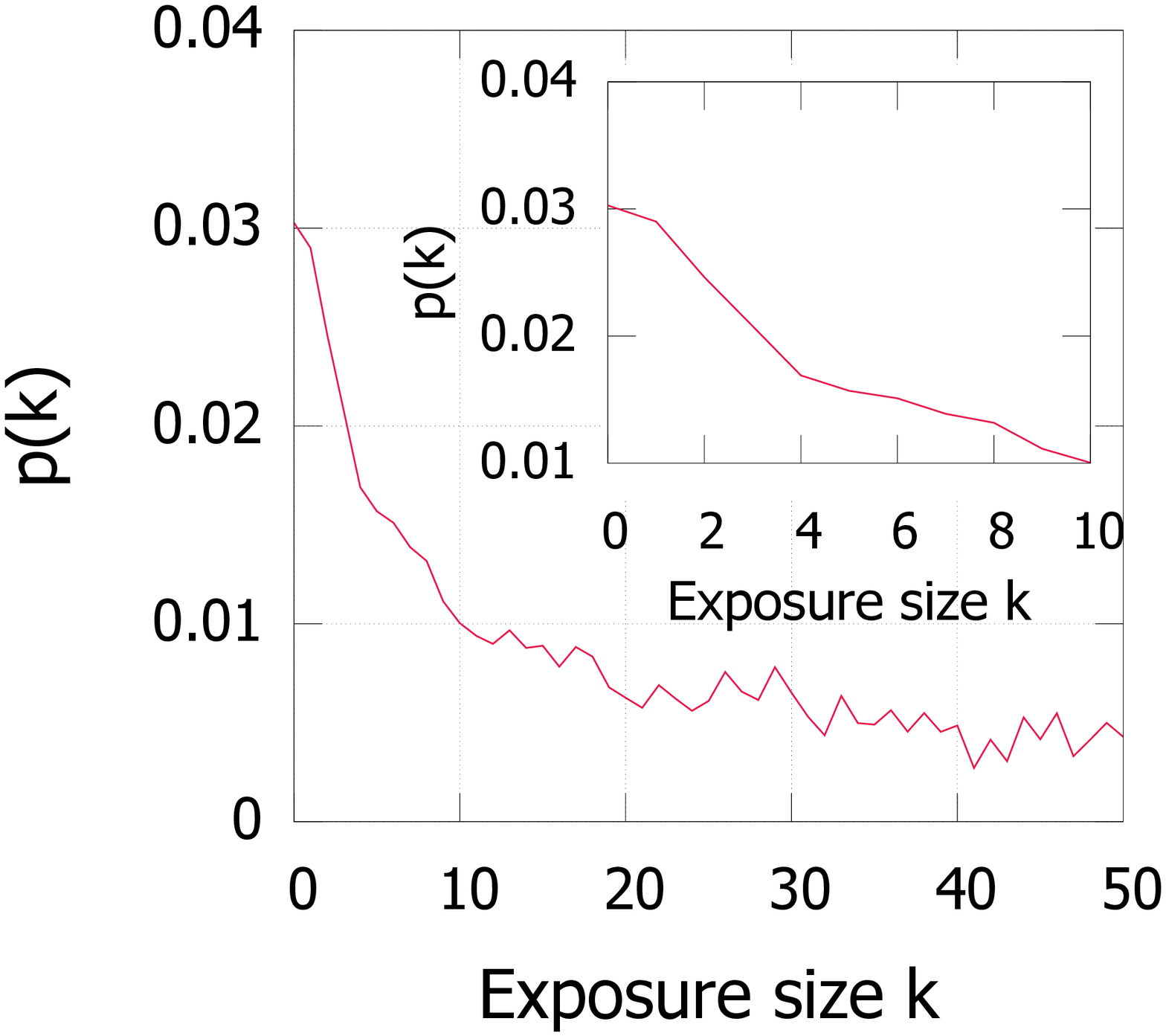}%
}
%\hspace{0.1in}
\subfloat[Lortab \label{fig:persistance_lortab}]{%
  \includegraphics[width=0.2\textwidth]{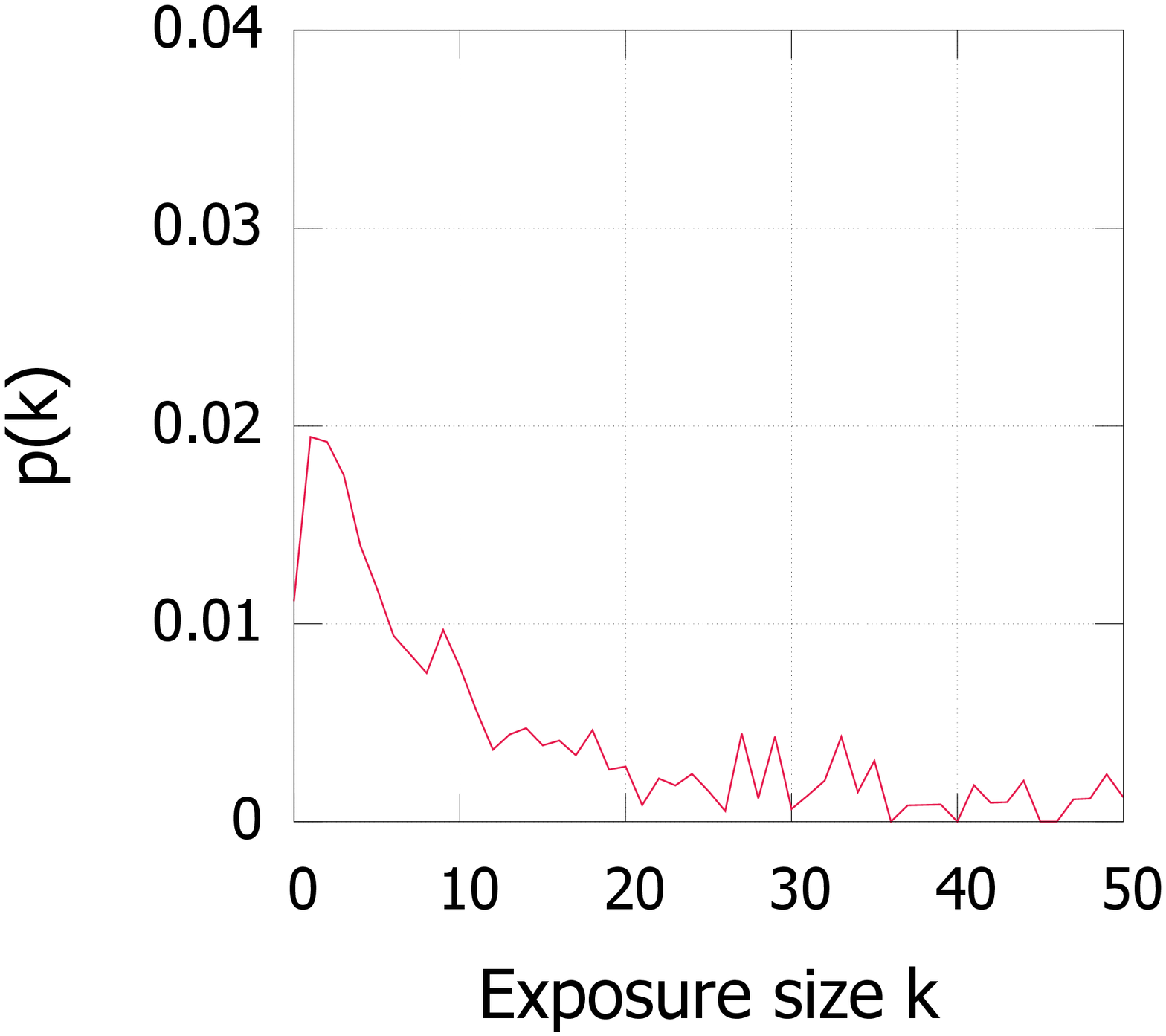}%
} 

\caption{Average exposure curve for drug names. $p(k)$ is the fraction of the network users who tweet about a particular drug-name directly after their $k^{th}$ exposure to it, given that they had not tweeted about it previously. The inset in figures~\protect\subref{fig:persistance_vicodin}, ~\protect\subref{fig:persistance_percocet} and ~\protect\subref{fig:persistance_oxycontin} shows the behavior of structural virality near the peaks at $k=1$.
}
%\vspace{-0.2in}
\label{fig:exposure}
\end{figure*}

\subsection{Role of Neighbors in Spreading} 
We next focus our attention on the influence of neighbors in the spreading of user engagement. For this experiment, cascades of size $> 20$ were considered. Table~\ref{tab:user_role_cascades} compares the properties of these cascades initiated by fringe nodes and non-fringe nodes. It is evident that cascades initiated by non-fringe nodes have a greater size on average with more nodes (aggregates for all cascades) in its first and second-hop. Irrespective of the type of the initiator, the presence of non-fringe nodes in the first-hop play an important role in inducting nodes in the second-hop of the cascades. This phenomenon is evident from cascades initiated by fringe nodes where only $5\%$ of first-hop nodes belong to the non-fringe category but bring in $45\%$ of the nodes in the second-hop.
It is also observed that the maximum width of the cascades initiated by fringe nodes occurs at more depth as compared to the ones generated by their counterparts. This further indicates that these cascades survive a few initial hops with the help of other fringe nodes only to peak later with the help of certain non-fringe ones, thus revealing an organic collaboration of the non-fringe as well as fringe nodes in the spreading process.

\par Thus we discover that all category of users, fringe and non-fringe as well as active and non-active, contribute significantly to generating the cascades,  highlighting a sizeable collective phenomenon. We next investigate the engagement behavior of users based on the contents (drug names) to discover whether Twitter serves as a more effective spreading media for certain types of drugs.

\begin{figure}[t]
\subfloat[Stickiness\label{fig:stickiness}]{%
  \includegraphics[width=0.5\linewidth]{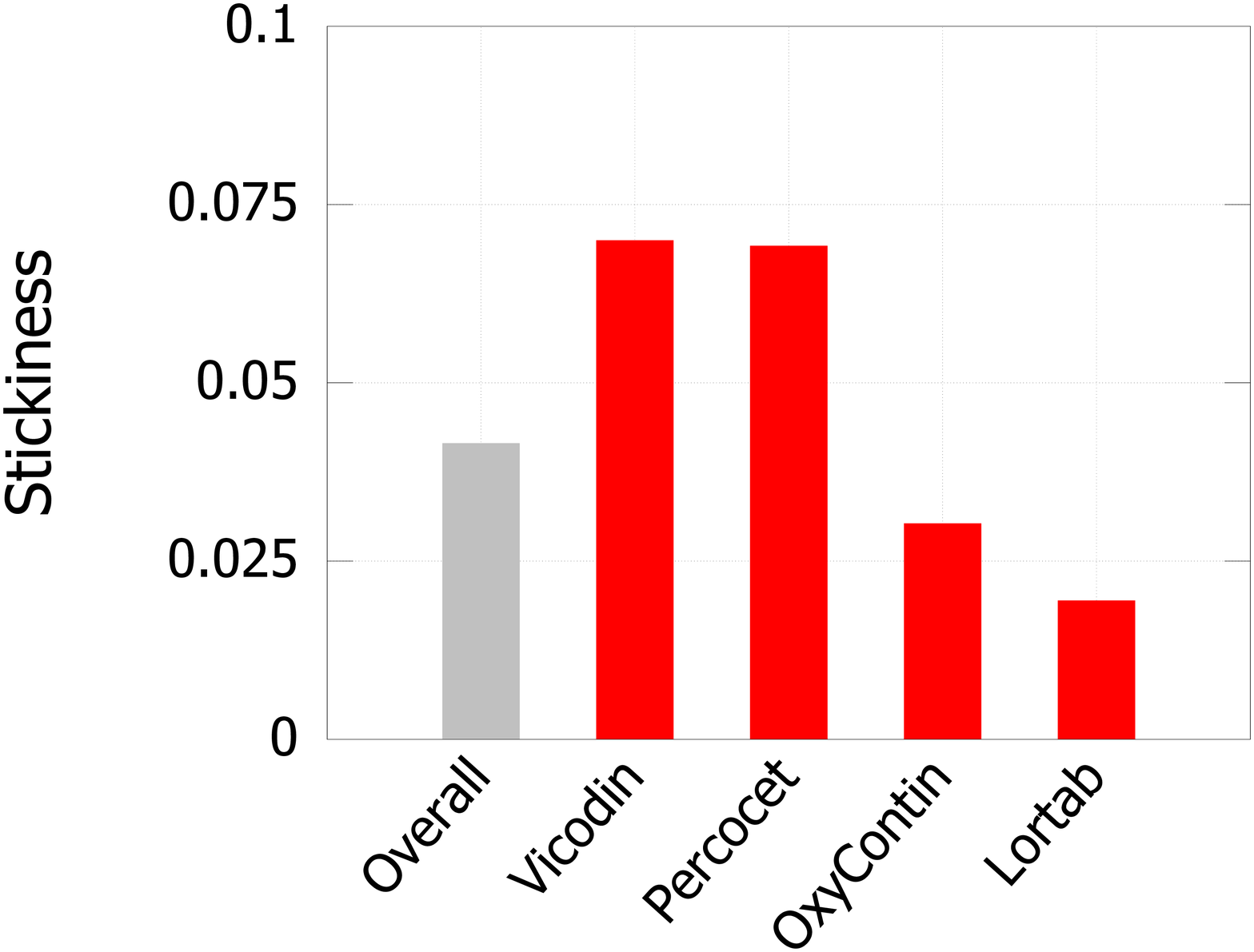}%
} 
\subfloat[Persistence\label{fig:persistance}]{%
  \includegraphics[width=0.5\linewidth]{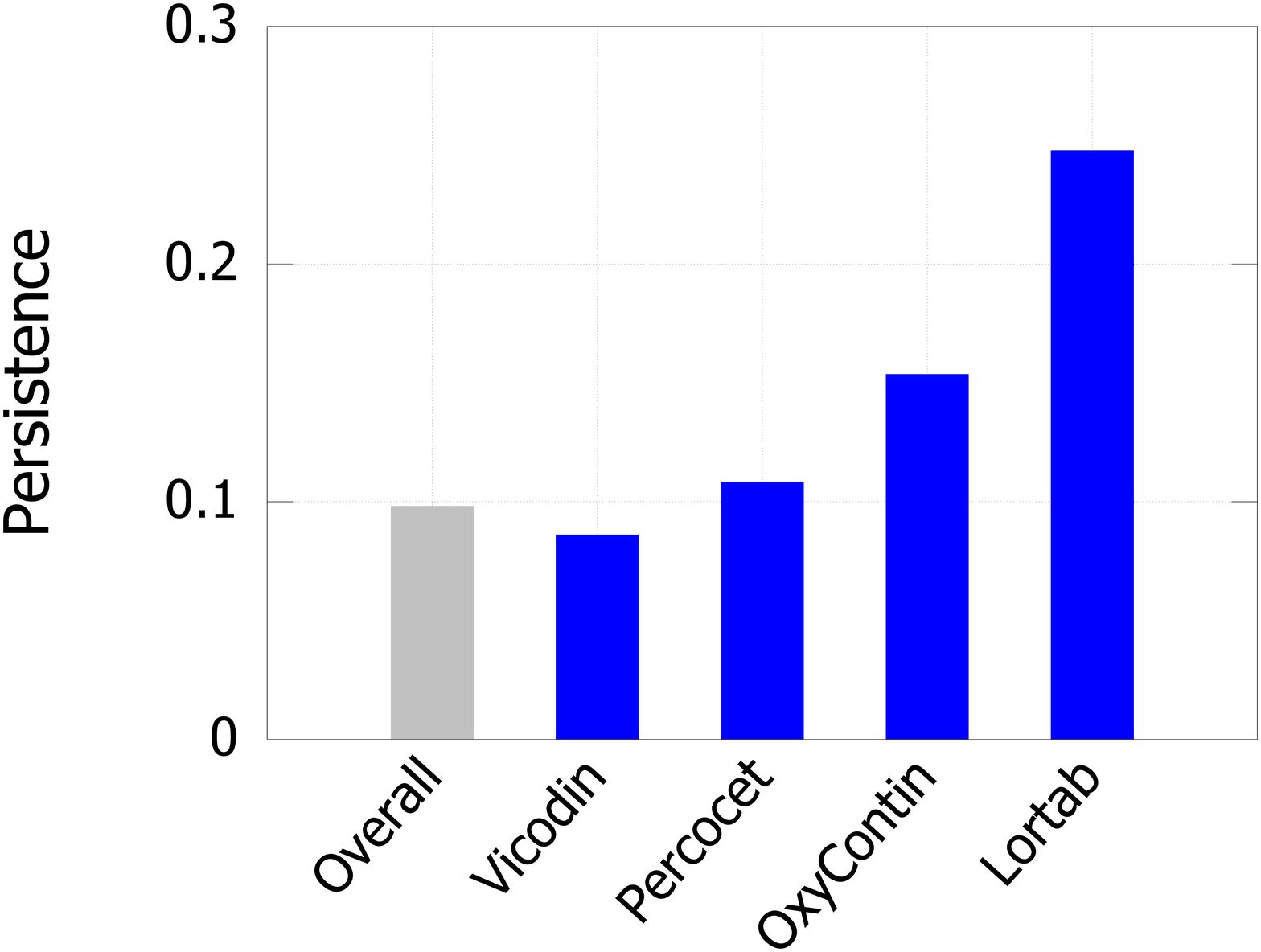}%
} 
\caption{Stickiness and persistence score measures for drug names as defined in ~\protect\cite{RomeroPersistenceWWW2011}.}
\label{fig:stickiness_persistance}
\end{figure}

\subsection{Stickiness and Persistence}
Information diffusion concerning different content types has been studied using measures like stickiness and persistence~\cite{spragueContagionPLOS2017,RomeroPersistenceWWW2011}. We use these measures to investigate the engagement behavior of the users with respect to the drug names. 
We investigate how repeated exposure to drug-abuse tweets with specific drug names influences the probability of adopting a similar engagement behavior. 
A user is considered to be $k$-exposed if there are $k$ users,   whom the current user follows, who have tweeted about drug-abuse. We use an ordinal time estimate measure for deriving the exposure curve $p(k)$, whereby, we calculate the number of users ($I(k)$) who generate their first drug-abuse tweet (an indication of adoption) after being $k$-exposed but before being $(k+1)$-exposed. This value is subsequently compared with the total number of $k$-exposed users ($E(k)$). The exposure curve is represented as $p(k)=\frac{I(k)}{E(k)}$. The stickiness is measured by the maximum value of $p(k)$ for all observed values of $k$ and the persistence $F(p)$ is represented by the ratio of the area under the exposure curve and the minimum area of the rectangle covering the exposure curve entirely. $F(p)$ provides a measure of the rate of decay in the adoption probability with an increasing number of exposures after it has reached the peak. A value of $F(p)$ near to $1$ indicates that repeated exposure to drug-abuse tweets would be required before the user herself starts engaging, indicating the presence of a complex contagion phenomenon.

\subsubsection*{Observations} 

We obtained the value of $p(k)$ for each drug type present in our dataset. 
Figure~\ref{fig:exposure} shows the average exposure curves for all the data and four major drug-names (determined based on \#exposures).  We observe in figure~\ref{fig:stickiness_persistance}\subref{fig:stickiness} Vicodin and Percocet, that were found to be mentioned in a significantly large number of tweets, have relatively much higher stickiness value ($0.0699$ and $0.0691$, respectively) compared to the other drugs. 
In both the cases, peaks are found at $k = 1$, indicating that users mostly engage themselves about these tweets after a single exposure only. A similar trend was observed for OxyContin, with a peak at $k = 0$.
This high value of stickiness is observed for these abused drugs due to their high popularity on Twitter.  
In contrast, Lortab has a relatively higher persistence of $0.24$ as seen in figure~\ref{fig:stickiness_persistance}\subref{fig:persistance}, hinting that repeated exposures continue to have marginal effects on user engagement. 
\par To explain the exceptionally high stickiness values for Vicodin (figure~\ref{fig:exposure}\subref{fig:persistance_vicodin}) and Percocet (figure~\ref{fig:exposure}\subref{fig:persistance_percocet}) at $k = 1$, we looked into the tweets containing these drug names. We observed that a significantly large number of tweets mentioning Vicodin and Percocet are related to the sale of these drugs (around $40,618$ and $38,910$ respectively). Since these tweets are generated independently, without being exposed, we see high values of $p(k)$ at $k=0$ and $1$ for Vicodin and Percocet. Further, since these drugs are popular among the drug-abusers, repeated exposures to tweets related to these drugs do not lead to any significant effect on user engagement, thus lowering the persistence. Users who are willing to discuss about these drugs rapidly engage themselves after one or two exposures. On the other hand, engagement for drugs, that are less popular over Twitter, shows high persistence. This could be possibly due to the fact that these drugs being less popular on Twitter, with increasing exposures the interest of the users about these drugs increases and hence the probability of engagement remains high with the number of exposures.

\section{Conclusion}
\label{sec:conclusion}
This paper provides a detailed analysis of the Twitter follower network involving around $0.42$ million users that are involved in the promotion of prescription drug-abuse using the Twitter platform and generating more than 50,000 cascades. We believe that this is a first major work involving such a large  scale of data that details the spreading of drug-abuse messages over Twitter. Analyzing the follower network of drug-abusers reveals a heavy core structure with high local connectivity among themselves, thereby providing various alternate channel of communication among the users. Investigations on the cascades of drug-abuse tweets helped us to discover certain major findings. It was discovered that the drug-abuse tweets spread over long paths across the Twitter follower network through groups of closely connected users in the network.
It was also observed that a significant percentage of cascades being initiated and driven by users with  low positional importance (with low count of followers as well as its followings), that we term as fringe nodes in the network. The spread over those cascades has been observed to be a result of a collective phenomenon involving both the important as well as the fringe nodes, indicating a resilience to  targeted elimination of few nodes. 
A diffusion model capturing these dynamics would be helpful in predicting drug related cascades. Considering the limited scope of this paper, we would like to develop such models as a possible extension of the current work.
Finally, observations suggest that drug-abusers on Twitter have much higher risk of adopting  newer drugs as increasing exposure of them enhances the probability of adoption. These findings necessitates a deeper and more detailed study of the abuse patterns and user behavior to control the spread of this menace.

\bibliographystyle{IEEEtran}
\bibliography{bibliography}

\end{document}